# Bayesian Convolutional Neural Network Based MRI Brain Extraction on Nonhuman Primates

**Gengyan Zhao**[a,*,1], **Fang Liu**[b,1], **Jonathan A. Oler**[c], **Mary E. Meyerand**[a,d], **Ned H. Kalin**[c], and **Rasmus M. Birn**[a,c]

[a]Department of Medical Physics, University of Wisconsin – Madison, USA

[b]Department of Radiology, University of Wisconsin – Madison, USA

[c]Department of Psychiatry, University of Wisconsin – Madison, USA

[d]Department of Biomedical Engineering, University of Wisconsin – Madison, USA

## Abstract

Brain extraction or skull stripping of magnetic resonance images (MRI) is an essential step in neuroimaging studies, the accuracy of which can severely affect subsequent image processing procedures. Current automatic brain extraction methods demonstrate good results on human brains, but are often far from satisfactory on nonhuman primates, which are a necessary part of neuroscience research. To overcome the challenges of brain extraction in nonhuman primates, we propose a fully-automated brain extraction pipeline combining deep Bayesian convolutional neural network (CNN) and fully connected three-dimensional (3D) conditional random field (CRF). The deep Bayesian CNN, Bayesian SegNet, is used as the core segmentation engine. As a probabilistic network, it is not only able to perform accurate high-resolution pixel-wise brain segmentation, but also capable of measuring the model uncertainty by Monte Carlo sampling with dropout in the testing stage. Then, fully connected 3D CRF is used to refine the probability result from Bayesian SegNet in the whole 3D context of the brain volume. The proposed method was evaluated with a manually brain-extracted dataset comprising T1w images of 100 nonhuman primates. Our method outperforms six popular publicly available brain extraction packages and three well-established deep learning based methods with a mean Dice coefficient of 0.985 and a mean average symmetric surface distance of 0.220 mm. A better performance against all the compared methods was verified by statistical tests (all p-values $< 10^{-4}$, two-sided, Bonferroni corrected). The maximum uncertainty of the model on nonhuman primate brain extraction has a mean value of 0.116 across all the 100 subjects. The behavior of the uncertainty was also studied, which shows the uncertainty increases as the training set size decreases, the number of inconsistent labels in the training set increases, or the inconsistency between the training set and the testing set increases.

*Corresponding Author: Gengyan Zhao, M.Sc., Department of Medical Physics, University of Wisconsin – Madison, 1111 Highland Avenue, Madison, Wisconsin 53705-2275., Phone: 608-772-7258, gzhao23@wisc.edu.
1Contributed equally.

# 1 Introduction

Brain extraction, also known as skull stripping, is an essential process in magnetic resonance imaging (MRI) and functional MRI (fMRI). It often serves as the first step in the preprocessing pipeline, since processing software often requires the extracted brains as sources and targets in the registration. By removing the non-brain parts, such as the skull, eyes, muscle, adipose tissue and layers of meninges etc., brain registration achieves improved performance (Wang et al., 2012). Meanwhile, the accuracy of brain extraction is important and can dramatically affect the accuracy of the following processes. Mistakenly removing brain tissues and/or retaining non-brain areas can lead to biased results of further analyses, such as the estimation of cortical thickness, parcel-wise averaged fMRI signal and voxel-based brain morphometry (Fennema-Notestine et al., 2006; Shattuck et al., 2009; van der Kouwe et al., 2008). Accurate brain extraction is extremely challenging as a result of complex brain anatomical structure, and therefore the improvement of brain extraction still remains an intensively investigated research topic (Roy et al., 2017).

Nonhuman primates have been widely used in neuroimaging as experimental subjects due to their similarity to human beings, especially in intervention studies and studies involving radiation, contrast agent and drugs (Baldwin et al., 1993; Kalin et al., 2007; Fox and Kalin, 2014). The particularity of the nonhuman primate's brain makes the challenge of brain extraction even more difficult. Nonhuman primates' brains are smaller in size than human brains, and have complex tissue structures. The eyes of nonhuman primates are relatively larger than human beings' and surrounded by much more adipose tissue. The adipose tissue behind their eyes are close to the brain, which makes it difficult to be separated. Their frontal lobes are quite narrow and protrude sharply (Rohlfing et al., 2012b), which causes this region to be excluded by many brain extraction packages. As a result, manually examining, refining or even extracting the whole brain is often unavoidable. Therefore, an accurate and robust automatic brain extraction approach for nonhuman primates is highly demanded to mitigate the time-consuming human intervention.

## 1.1 Previous Work

A large number of brain extraction methods have been proposed in recent decades, which again emphasize its importance. However, the need for an accurate, robust and sufficiently fast method has not yet been fulfilled. Mainly, these methods can be divided into two categories, edge based methods and template based methods (Roy et al., 2017). Although most of these methods work well for human brains, they encounter challenges when dealing with nonhuman primate brains due to their complex anatomical structure (Wang et al., 2014). A comprehensive review can be found at (Roy et al., 2017; Wang et al., 2014). Due to the sub-optimal performance of existing automated brain extraction routines in rhesus monkeys, prior work from our laboratory has used brain images that were manually extracted by well-trained experts (Oler et al., 2010; Birn et al., 2014; Fox et al., 2015a). This procedure, however, is extremely time consuming and labor intensive. In this study, we develop a new automated method based on deep learning, and compare this new method to six state-of-the-art automated brain extraction packages including both edge-based and template-based methods, and three well-established deep learning based methods. These tools are commonly used and are all freely available.

The Brain Extraction Tool (BET) (Smith, 2002) is based on a deformable tool, which initializes a spherical mesh at the center of gravity of the brain, and then expands it towards the edge of the brain. The whole process is guided by a set of locally adaptive forces determined by surface smoothness and contrast changes in the vicinity of the surface. This toolbox has been reported to be useful for nonhuman primate brain extraction (https://fsl.fmrib.ox.ac.uk/fsl/fslwiki/BET/FAQ).

The Brain Surface Extractor (BSE) was proposed by (Shattuck et al., 2001). It involves anisotropic filtering, Marr-Hildreth edge detection and morphological operations. Serving as an edge-preserving filter, the anisotropic diffusion filtering step is intended to smooth gentle gradients, while preserving sharp gradients, which helps the edge detection. Morphological operations, including erosion and dilation, can further enhance the results from edge detection.

Another popular brain extraction tool is 3dSkullStrip in AFNI (https://afni.nimh.nih.gov/afni/). It consists of 3 steps: removing gross non-uniformity artifacts, iteratively expanding a spherical surface to the edge of the brain and creating masks and surfaces of the brain. The second step is a modified version of BET. The improvement includes excluding eyes and ventricles, driving the expansion with data both inside and outside of the surface, and involving three-dimensional (3D) edge detection. It has a -monkey option helping with initialization of a surface based on nonhuman primate brains.

The Hybrid Watershed Algorithm (HWA) (Ségonne et al., 2004) combines watershed algorithm with a deformable surface model. Based on the 3D white matter connectivity, the watershed algorithm outputs the initial estimate of the brain volume, and then the deformable model generates a force field to drive a spherical surface to the boundary of the brain. The option -atlas can incorporate a statistical brain atlas generated from accurately segmented human brains to correct the segmentation. The HWA in the software FreeSurfer is not originally designed or optimized for nonhuman primates.

Robust Learning-Based Brain Extraction (ROBEX) (Iglesias et al., 2011) is a more recently published algorithm. In ROBEX a discriminative method is combined with a generative model. A random forest classifier is trained to detect the contour of the brain, after the subject is registered to the template with an affine transformation, and then a point distribution model is fitted to find the most likely boundary of the brain. Finally, a small deformation optimized by graph cuts serves as the refining step.

Recently the National Institute of Mental Health Macaque Template (NMT) was published by (Seidlitz et al., 2017). It is a high-resolution template for the macaque brain derived from MRI images averaged from 31 subjects. Rhesus macaque brain MRI images can be registered into this template space to extract the brains. After brain extraction in the template space, the brain images can be transformed back to the original image space with the inverse transformation. To increase the accuracy of the registration, both affine and deformable transformations should be involved.

In the recent past, tremendous progress has been made in deep learning as a result of the revival of deep neural networks (Krizhevsky et al., 2012a; LeCun et al., 2015) as well as the rapid advance of parallel computing techniques (Coates et al., 2013; Schmidhuber, 2015). Among many deep learning methods, deep convolutional neural networks (CNN) have proven to be very useful in a broad range of computer vision applications, outperforming traditional state-of-the-art methods (Shelhamer et al., 2016; Simonyan and Zisserman, 2014). In the area of image segmentation, many CNN based architectures have been proposed. These methods can broadly be divided into two-dimensional (2D) methods and 3D methods. A fully convolutional network (FCN) was proposed by (Long et al., 2015) as a 2D network for the general task of semantic segmentation. SegNet (Badrinarayanan et al., 2015b) was first proposed for road and indoor scene segmentation, and later was combined with a 3D deformable model to solve tissue segmentation in MRI (Liu et al., 2017). UNet (Ronneberger et al., 2015) is a kind of 2D encoder-decoder network proposed for microscopic images, and later expanded to 3D for volumetric data (Çiçek et al., 2016). VNet (Milletari et al., 2016) was proposed as a 3D FCN with dice loss to perform 3D segmentation on MR images. (Wachinger et al., 2017) proposed a 3D patch-based method and arranged 2 networks hierarchically to separate the foreground and then identify 25 brain structures. (Chen et al., 2017) proposed a 3D residual network with multi-modality and multi-level information to identify 3 key structures of the brain. Recently, Several patch-based 3D FCN were also proposed, like LiviaNet to segment the subcortical region of the brain (Dolz et al., 2017) and DeepMedic to segment brain lesions (Kamnitsas et al., 2017). In the specific field of brain extraction, (Kleesiek et al., 2016) proposed a 3D patch based CNN network for human brain extraction on T1w human brain datasets and a multi-modality human brain dataset with tumors. In a further study, (Salehi et al., 2017) proposed an auto-context CNN where the probability maps output by the CNN are iteratively used as input to the CNN along with the original 2D image patches to refine the results. A more comprehensive review can be found at (Bernal et al., 2017; Craddock et al., 2017). In this study, we also compared the proposed method to three deep learning based methods, SegNet, LiviaNet and VNet. They are 2D, patch-based 3D and 3D networks respectively.

## 1.2 Our Approach

The purpose of this work is to implement and validate deep learning based methods on nonhuman primate brain extraction, and to build a framework for this fully automatic approach. In this study, we propose to improve the brain extraction accuracy using a Bayesian CNN with refinement through fully connected 3D conditional random field (CRF). In comparison to previous brain extraction studies, our study has several novel aspects. Firstly, we evaluated brain extraction using Bayesian SegNet, a Bayesian convolutional encoder-decoder network that involves Monte Carlo dropout layers to provide additional information for model uncertainty evaluation. In our previous study, the basic version of this network, SegNet, was proven to be highly efficient in MRI tissue segmentation (Liu et al., 2017). Secondly, we incorporated the fully connected 3D CRF as a post-processing step to regularize the result according to the fully 3D anatomical context. Fully connected 3D CRF, as a probabilistic graphic model, is helpful to improve 2D CNN segmentation results by considering the distance and contrast relationships among all the voxel pairs in the whole 3D space. Finally, on a large-scale nonhuman primate dataset, we made a full comparison of the proposed method with current state-of-the-art software packages and well-established deep learning based methods. The accuracy and robustness of these algorithms on challenging nonhuman primate brain extraction were investigated. We hypothesize that a Bayesian deep learning based image segmentation framework with fully connected 3D CRF refinement is suitable for nonhuman primate's brain extraction with improved accuracy and efficiency, and the uncertainty it generated can also reflect the confidence of the model on each prediction.

## 2 Material and Methods

### 2.1 Image Datasets

MRI data of 100 periadolescent rhesus macaques (Macaca mulatta; mean (standard deviation) age = 1.95 (.38) years; 43% female) were collected in a 3T MRI scanner (MR750, GE Healthcare, Waukesha, WI, USA) with a 16-cm quadrature birdcage extremity coil (GE Healthcare, Waukesha, WI, USA) and a stereotactic head-frame integrated with the coil to prevent motion. Immediately prior to the scan, subjects received medetomidine (30 μg/kg i.m.) and a small dose of ketamine (<15 mg/kg) for anesthesia purpose. During the scan, anatomical structures were acquired using a 3D T1-weighted inversion-recovery fast gradient echo sequence with the following imaging parameters: TE = 5.41ms, TR = 11.39, TI = 600ms, Flip Angle = 10°, NEX = 2, FOV = 140 mm, Bandwidth = 61.1 kHz. The whole brain was reconstructed into a 3D volume of 256×224 in-plane matrix size and 0.27×0.27 mm$^2$ in-plane pixel size with 248 slices over 124 mm. All the brains were then manually extracted by well-trained image scientists using these T1 images with the software SPAMALIZE (http://psyphz.psych.wisc.edu/~oakes/spam/spam_frames.htm). The data used in this study are a subset of those used in our prior studies (Fox et al., 2015b; Shackman et al., 2017).

### 2.2 Full Brain Extraction Method

The proposed brain extraction pipeline is a combination of Bayesian SegNet (Kendall et al., 2015a) and fully connected 3D CRF (Krähenbühl and Koltun, 2012). As shown in Fig. 1 It has a training phase and a testing phase. The 3D brain image volumes and corresponding manual label volumes are treated as a stack of 2D images input to the Bayesian neural network. In the training phase, the process is formulated as an optimization problem to optimize the network parameters by minimizing the difference between the network's output and the manual labels using multinomial logistic loss (Krizhevsky et al., 2012b). In the testing stage, the network with well-trained parameters are used as a pixel-wise segmentation classifier to predict the label probability and generate model uncertainty on each pixel of new brain volumes. Finally, the predicted probabilities and the 3D brain volumes are passed to fully connected 3D CRF for refinement in the whole 3D context.

## 2.3 Bayesian Convolutional Neural Network

A convolutional encoder-decoder network, Bayesian SegNet, is used as the core segmentation engine in the brain extraction workflow (Fig. 1). This network was first introduced by (Kendall et al., 2015a) and benchmarked on the multiple scene recognition datasets (Everingham et al., 2015) with excellent performance. This network consists of a VGG16 (Simonyan and Zisserman, 2014) encoder network and a reversed decoder network. The encoder network performs the function of feature extraction and data compression, while the decoder network assembles the compressed features to the original image size using extracted features via multi-scale sparse upsampling (Badrinarayanan et al., 2015b). Networks of encoders and decoders were constructed using a series of convolutional layers, batch normalization (Ioffe and Szegedy, 2015), ReLU non-linearity (Nair and Hinton, 2010), and maximum pooling layers or upsampling layers. Compared with other segmentation CNNs, Bayesian SegNet features both dropout training and dropout testing. Dropout training offers the network robustness against overfitting especially on small datasets. Dropout testing predicts both pixel-wise probability maps for all the labels as well as additional measurement of model uncertainty which is particularly useful for accuracy evaluation. These 2 features are achieved by implementing Bayesian SegNet with Monte Carlo dropout layers as shown in Fig. 1. The dropout rate is set beforehand and a certain percentage of neurons in the preceding layer are randomly ignored in every iteration during training or every forward pass during testing (Srivastava et al., 2014).

Given the dataset $X$ and its corresponding label set $Y$, (Gal and Ghahramani, 2015) showed that Monte Carlo dropout training can be used to evaluate the posterior distribution over the network weights $W$:

$$p(W \mid X, Y) \quad (1)$$

Since this posterior is not traceable directly from Bayesian SegNet, an approximation can be made by using variational inference (Gal and Ghahramani, 2015; Kendall et al., 2015a), which allows defining an approximating distribution $q(W)$ and inferring it by minimizing the Kullback-Leibler (KL) divergence (Gal and Ghahramani, 2015):

$$KL(q(W) \parallel p(W \mid X, Y)) \quad (2)$$

(Gal and Ghahramani, 2015) illustrated that the integral in the KL divergence can be approximated with Monte Carlo integration over the network weights, and the process of minimizing the KL divergence is equivalent to performing Monte Carlo dropout training.

(Gal and Ghahramani, 2015) also showed that after getting the optimal weights, Monte Carlo dropout sampling can also be used in testing. To predict the label $y^*$ for the data $x^*$, the posterior distribution can be determined through $T$ times Monte Carlo dropout testing. During each testing the network weight subset $\hat{W}_t$ is occupied.

$$p(y^* \Big| x^*, X, Y) \approx \int p(y^* \mid x^*, W) q(W) dW \approx \frac{1}{T} \sum_{t=1}^{T} p(y^* \mid x^*, \widehat{W}_t) \tag{3}$$

$$\widehat{W}_t \sim q(W)$$

The integral in the equation is approximated with Monte Carlo integration, which is identical to Monte Carlo dropout sampling of Bayesian SegNet during testing. This can be considered as sampling the posterior distribution over the weights to get the posterior distribution of the predicted label probabilities. The mean of sampled probabilities $p(y^* \mid x^*, \widehat{W}_t)$ will be used as the prediction of the probability map for each label, and the variance of them will be used as the model uncertainty on each prediction.

## 2.4 Fully Connected Three-Dimensional Conditional Random Field

The final prediction outputs from the Bayesian SegNet are 2D probability maps for each label. To take into account the 3D contextual relationships among voxels, we propose to incorporate fully connected 3D conditional random field (CRF) (Krähenbühl and Koltun, 2012) to refine the results from the Bayesian SegNet. Based on the probability maps from Bayesian SegNet, this approach can maximize the label agreement between voxels having similar contrasts or close to each other in the whole 3D volume by a maximum a posteriori (MAP) inference (He et al., 2004) made in the CRF defined over the full brain volume. Considering $x$ as the label assignment for each voxel, and, $i, j$ as the voxel index ranging from 1 to the total number of voxels, to get the MAP inference optimization is carried out to minimize the Gibbs energy in the 3D space:

$$E(x) = \sum_i \psi_u(x_i) + \sum_{i<j} \psi_p(x_i, x_j) \tag{4}$$

The probability result on each voxel from the Bayesian SegNet is used to build the unary potential $\psi_u(x_i)$, while the pairwise potential $\psi_p(x_i, x_j)$ depends on each voxel pair's location $p_i, p_j$ and intensity $I_i, I_j$:

$$\psi_p(x_i, x_j) = \mu(x_i, x_j) \left[ \omega_1 \exp\left( -\frac{\|p_i - p_j\|^2}{2\theta_\alpha^2} - \frac{|I_i - I_j|^2}{2\theta_\beta^2} \right) + \omega_2 \exp\left( -\frac{\|p_i - p_j\|^2}{2\theta_\gamma^2} \right) \right] \tag{5}$$

In the pairwise potential, the appearance kernel and the smoothness kernel are involved (Krähenbühl and Koltun, 2012). The appearance kernel, the first exponential term in Eq. 5, assumes voxels close to each other or having similar contrasts tend to share the same label. The extent of each effect is controlled by $\theta_\alpha$ or $\theta_\beta$. The smoothness kernel, the second exponential term, removes isolated small regions (Shotton et al., 2009), and is controlled by $\theta_r$. $\omega_1$ and $\omega_2$ are the weights for the two kernels. The compatibility function, $\mu(x_i, x_j)$, is set as the Potts model:

$$\mu(x_i, x_j) = 1_{[x_i \neq x_j]} \quad (6)$$

To make the complex inference practical given a tremendous number of pairwise potentials in fully connected CRF, we use the highly efficient algorithm proposed by (Krähenbühl and Koltun, 2012), where the pairwise edge potentials are defined as a linear combination of Gaussian Kernels in the feature space. A mean approximation to the CRF distribution is made in the algorithm, and it is optimized through an iterative message passing process. (Krähenbühl and Koltun, 2012) showed that the message passing process can be performed using Gaussian filtering in the feature space. In this way, using highly efficient approximations of high-dimensional filtering, the computational complexity of message passing can be reduced from being quadratic to being linear, with respect to the number of variables. As a result, the approximate inference algorithm for fully connected 3D CRF is linear with respect to the number of variables and sublinear with respect to the number of edges in the model.

## 2.5 Parameter Selection for Competing Methods

The proposed method was compared to six popular publicly available brain extraction software packages and three state-of-the-art deep learning based methods, including 3dSkullStrip in AFNI (17.0.09; https://afni.nimh.nih.gov/), BET (Smith, 2002) in FSL (5.0.10; https://fsl.fmrib.ox.ac.uk/fsl/fslwiki), BSE (Shattuck et al., 2001) in BrainSuite (v. 17a; http://brainsuite.org/), HWA (Ségonne et al., 2004) in FreeSurfer (Stable v6.0; https://surfer.nmr.mgh.harvard.edu/), ROBEX (1.2; https://www.nitrc.org/projects/robex) (Iglesias et al., 2011), NMT (v1.2; https://github.com/jms290/NMT) (Seidlitz et al., 2017), SegNet (https://github.com/alexgkendall/caffe-segnet) (Badrinarayanan et al., 2015c), LiviaNet (https://github.com/josedolz/LiviaNET) (Dolz et al., 2017) and VNet (https://github.com/faustomilletari/VNet) (Milletari et al., 2016). For a direct comparison, SegNet used the same parameters as Bayesian SegNet. LiviaNet used all the default parameters (30 epochs; 20 subepochs per epoch; 1000 samples in each subepoch). VNet used the default parameters, 5000 iterations (500 epochs) and batch size 1 due to the limitation of GPU memory, and the learning rate was changed to 0.00015 accordingly. To determine the parameters of the other software packages, a two-step evaluation strategy was used for each software package, and the parameter selection was done under the assumption that all the subjects are similar enough to one another that they can be properly processed with one set of parameters. We first randomly chose one representative subject, varied each parameter by small increments in either direction from the default values to achieve the best accuracy by careful visual inspection by a well-trained researcher. The selected parameters were then tested on a second randomly selected subject to verify its validity before application to the rest of the dataset. In our experiments, no obvious difference was observed on the performance of the selected set of parameters between the 2 selected subjects. With the exception of ROBEX and NMT, which do not have any parameters to tune, all the other software packages' parameters studied and their associated values used are shown in Table 1. For 3dSkullStrip in AFNI, besides the original method labeled as AFNI in this paper, to achieve a better performance in detecting the protruding frontal lobe, we also performed 3dSkullStrip with a coronal slice thickness reduced to one half of the original value in the headers of the data. This method is labeled as AFNI+ later in this paper. This was done because a common challenge in skull stripping rhesus macaque brains is that the ventral portion of the frontal lobe is quite narrow in the transverse plane, and the resultant high curvature of this region causes many skull stripping algorithms to exclude the anterior portions of the frontal lobe. By reducing the coronal slice thickness by a factor of 2, the curvature is reduced, and the frontal lobe is more easily retained. The slice thickness is then set back to the original value after brain extraction. Note that this procedure does not involve any resampling; only the value of the slice thickness in mm is changed. For the NMT, the AFNI functions, align_epi_anat.py and auto_warp.py (https://afni.nimh.nih.gov/) are used to carry out the 12 degrees of freedom (DOF) affine and deformable registration between the original image and the template for brain extraction in the NMT template space.

## 2.6 Metrics for Comparison

Several quantitative metrics commonly used in image segmentation were used to evaluate the performance of all the compared brain extraction methods (Kleesiek et al., 2016; Taha and Hanbury, 2015; Wang et al., 2014). Let M and R represent the brain mask extracted by a specific method and the manually extracted reference serving as the ground truth respectively, then the following metrics can be defined: True Positive: $TP = M \cap R$; True Negative: $TN = \bar{M} \cap \bar{R}$; False Positive: $FP = M \cap \bar{R}$; False Negative: $FN = \bar{M} \cap R$; Sensitivity: $Sens = \frac{TP}{TP + FN}$; Specificity: $Spec = \frac{TN}{TN + FP}$; Absolute Error: $E_{obs} = FP \cup FN$. We also involved the most commonly used metrics in image segmentation, dice coefficient (DC) (Dice, 1945), maximum symmetric surface distance (or Hausdorff distance, HD) (Huttenlocher et al., 1993) and average symmetric surface distance (ASSD) (Geremia et al., 2011):

$$DC = \frac{2|M \cap R|}{|M| + |R|} = \frac{2TP}{2TP + FP + FN} \quad (7)$$

$$HD = \max\left( \max_{m \in \partial(M)} \min_{r \in \partial(R)} \left\| m - r \right\|, \max_{r \in \partial(R)} \min_{m \in \partial(M)} \left\| r - m \right\| \right) \quad (8)$$

$$ASSD = \frac{\sum_{m \in \partial(M)} \min_{r \in \partial(R)} \left\| m - r \right\| + \sum_{r \in \partial(R)} \min_{m \in \partial(M)} \left\| r - m \right\|}{|\partial(M)| + |\partial(R)|} \quad (9)$$

Where ‖ ‖ means the total voxels in the set, and $\partial( )$ means the boundary of the set. The Dice coefficient is probably the most widely used metric for image segmentation. It takes the real value within [0,1], where 1 means a perfect segmentation, and 0 means there is no overlap at all. For the segmentation of a region as large as the brain, the Dice coefficient is less sensitive due to the small edge to volume ratio. Hence, we also introduced the surface distance based metrics. HD is defined as the maximum shortest Euclidean distance between two surface sets, while ASSD is defined as the average of these shortest Euclidean distances. HD or ASSD is 0 for a perfect segmentation. Both of these are used as segmentation metrics historically, but since HD is sensitive to outliers, ASSD is usually preferred (Gerig et al., 2001; Zhang and Lu, 2004).

False positive, false negative, and absolute error maps are all spatial error maps. To visualize the systematic spatial error distribution of each method, the averaged error maps are calculated. First, the 12-DOF affine registration and deformable registration are done for each subject's full brain image from the original space to the NMT (Seidlitz et al., 2017) space using AFNI's align_epi_anat.py and auto_warp.py (https://afni.nimh.nih.gov/). Then, each kind of error map for each method are transformed to the NMT space with the transformation matrices calculated in the first step. Next, each specific kind of error map is averaged across all the subjects in the NMT space. Finally, for display purposes, the natural logarithm of the averaged error maps collapsed (averaged) along each axis was plotted (Kleesiek et al., 2016; Wang et al., 2014).

## 2.7 Experiments

### 2.7.1 Brain Extraction for Nonhuman Primates—Before being sent to Bayesian SegNet as inputs, each subject's original 3D image volume was normalized to [0,1] and disembled along the longitudinal (superior-inferior) axis into a stack of 2D images. Since manual skull stripping was done on the 3D brain volumes that were manually cropped to exclude the body of the monkey and regions far outside the brain, all the 2D images and corresponding manual labels were upsampled to the same size (352×256) with bilinear interpolation and nearest neighbor interpolation respectively. The Bayesian SegNet was trained by Stochastic Gradient Descent (SGD) algorithm with multinomial logistic loss in 60000 iterations (18 epochs). The learning rate during training was fixed as 0.01 with a momentum of 0.9. In testing, the number of samples were set as 6 based on (Kendall et al., 2015a) and the limitation of GPU memory. The dropout rate for all the dropout layers were set as 0.5 for both MC dropout training and testing (Kendall et al., 2015a). Fully connected 3D CRF was performed for each subject following Bayesian SegNet. The parameters for fully connected 3D CRF were empirically selected in the same manner as was described in Section 2.5: $\omega_1 = 3$, $\omega_2 = 1$, $\theta_\alpha = \theta_\gamma = 4$ and $\theta_\beta = 1$ (Eq. 5), and a total of 5 iterations were carried out to refine each subject's result. The whole processing pipeline is implemented on the platform of Caffe (Jia et al., 2014) based on the original work of (Kendall et al., 2015; Krähenbühl and Koltun, 2012). The 100-subject dataset was divided into 2 sets by random permutation, resulting in 50 subjects in each half. A two-fold cross-validation was performed between these 2 sets to test the proposed method on all the subjects. In this way, the training and testing phases used independent sets of data. Due to the robustness of deep learning based methods, no registration is used during the entire process. All the training and testing of the proposed method and the evaluation of other compared methods was performed on a workstation hosting 2 Intel Xeon(R) E5-2620 v4 CPUs (8 cores, 16 threads @2.10GHz) with 64 GB DDR4 RAM and a Nvidia GTX980Ti GPU with 6 GB GPU memory. The workstation runs a 64-bit Linux operation system.

**2.7.2 Uncertainty of the Bayesian SegNet—**One important aspect of Bayesian SegNet is the output of model uncertainty on predictions. We studied the influence of training set size, label inconsistency and training-testing inconsistency on the model uncertainty. Besides the training set of 50 subjects, we trained the Bayesian SegNet with 25 and 5 subjects to show how the size of the training set can affect the uncertainty. We also trained Bayesian SegNet with 50 subjects, of which either 25 or 5 subjects had sub-optimal labels from AFNI +, while the rest of the labels were the manually-segmented ground truth. This was done to investigate how label consistency affects the uncertainty. Since AFNI+ usually includes some non-brain tissues around the frontal lobe and includes the adipose tissue behind the eyes in nonhuman primates, these labels can be used to simulate the same kind of errors possibly made in manual labels by carelessness or fatigue. To achieve similar level of convergence, the training procedures in this section were also performed with 60000 iterations. Finally, to study the influence of inconsistency between the training set and the testing set, 4 new testing sets were designed to be slightly inconsistent with the training set. The 50 subjects in fold 2 (set #2) were rotated around the longitudinal axis by 10, 20 and 30 degrees, to create 3 new testing sets. Another 50 subjects (mean (standard deviation) age = 3.20 (.86) years; 66% female) were scanned at another site, with an older scanner model (GE Signa 3T, Waukesha, WI, USA), but the same coil model and experimental setup. The brain masks for these 4 new testing sets were all generated by the Bayesian SegNet trained with the original 50 subjects in fold 1 (set #1) (mean (standard deviation) age = 1.98 (.37) years; 38% female) and processed by 3D CRF with the same parameters used in Section 2.7.1. These results were then compared to the results of the original 50 subjects in fold 2 (mean (standard deviation) age = 1.92 (.39) years; 48% female) tested in Section 2.7.1. Model uncertainties on these new testing sets were also generated and compared with those on the original fold 2 in Section 2.7.1.

# 3 Results

## 3.1 Convergence of Bayesian SegNet during Training

Fig. 2 shows the convergence of the Bayesian SegNet on a 50-subject nonhuman primate training set and the convergence speed in one training set. There is no obvious improvement in the loss and accuracy after 18 epochs. Without loss of generality, we used the network weights at the 18th epoch, which is equivalent to 60000 iterations to predict brain masks, and one 18-epoch training over 50 subjects took about 12.3 hours on our workstation.

## 3.2 Brain Extraction for Nonhuman Primates

The performance of the proposed method and nine other state-of-the-art-methods were evaluated on the T1 volumes from 100 subjects. The Dice coefficient and average symmetric surface distance of each individual for each method are plotted in Fig. 3. The boxplots are shown in Fig. 4, and the corresponding mean values and standard deviations are shown in Table 2. Fig. 3 shows that the performance of the proposed method, a combination of Bayesian SegNet and fully connected 3D CRF (BSegNetCRF), is the best on both metrics among all the compared methods for each individual's brain extraction. The boxplots in Fig 4 shows the median and quartiles of the Dice coefficient and the average symmetric surface distance for each method. Table 2 illustrates that BSegNetCRF has not only achieved the best mean values, but also the smallest standard deviation on both metrics. Multiple pairwise Wilcoxon signed rank tests (two-sided) were done to compare the performance of these methods. The performance of BSegNetCRF is better than all other methods, as evaluated on both metrics ($p < 10^{-4}$, Bonferroni corrected). BSegNet is significantly better than SegNet on both metrics ($p < 10^{-4}$, Bonferroni corrected). In the comparison between BSegNet and VNet, VNet's average symmetric surface distance is significantly better than BSegNet's ($p < 10^{-4}$, Bonferroni corrected) but the p-value on Dice Coefficient is 0.0425 before Bonferroni correction, which is insignificant after Bonferroni correction at the 0.05 significance level. Both BSegNet and VNet are better than LiviaNet on both metrics ($p < 10^{-4}$, Bonferroni corrected). The comparisons of different methods on Hausdorff distance, sensitivity and specificity are also shown in Supplementary Fig. S1-S4.

Fig. 5 shows the extracted brain masks from all the methods for a representative subject. AFNI typically cannot catch the complete frontal lobe due to its challenging sharp curvature in nonhuman primates. AFNI+ was used to fix this by reducing the coronal slice thickness to one half. Although AFNI+ can capture the frontal lobe more completely, but it also captures tissues outside the brain. In addition, both AFNI and AFNI+ mistakenly include the adipose tissue behind the eyes. BET misses the frontal and occipital lobes of the brain, and the mask often extends past the upper boundary of the brain. BSE and HWA are not designed for nonhuman primates, and their resultant brain masks include a lot of non-brain tissue. ROBEX failed on all the nonhuman primate data, so it is not shown in the figure. NMT mistakenly includes the adipose tissue behind the eyes as part of the brain mask, and misses some boundaries and overshoots some others. SegNet tends to include the nonbrain tissue around the frontal lobe, eyes and brain stem as part of the brain mask. LiviaNet includes multiple nonbrain regions and misses some small regions within the brain. Results from BSegNet, VNet and BSegNetCRF are very close to the manually labeled ground truth. VNet performs well at the frontal lobe and eyes, but in general it includes slightly more nonbrain voxels close to the boundaries than BSegNetCRF, especially at the area close to the bottom of the brain and the brain stem. BSegNetCRF is also better than BSegNet, especially at excluding the brain stem. Fig. 5 shows a comparison of the error maps of these methods on the representative subject. A more comprehensive systematic comparison is shown in Fig. 6.

Fig. 6 is the averaged absolute error map of each method in the NMT template space. As mentioned in Section 2.6, the natural logarithm of the averaged error maps collapsed (averaged) along each axis is shown for display purposes. In Fig. 6's comparison, BSegNetCRF has the best systematic performance with a much smaller error distribution than other methods considering the results in all the voxels for every subject, and the systematic performance improvement by fully connected 3D CRF can also be viewed between the absolute error maps of BSegNet and BSegNetCRF. VNet also has very good performance, but BSegNetCRF is still better than VNet around the bottom area of the brain. The averaged false positive and false negative maps can also be found in the Supplementary Fig. S5 and S6.

As a probabilistic network, Bayesian SegNet is able to output the model uncertainty on the prediction of every voxel's label via Monte Carlo dropout testing. The maximum voxel-labeling uncertainty of Bayesian SegNet within each subject's 3D volume was calculated, and it has a mean value of 0.116 and a standard deviation of 0.023 across all 100 subjects. Fig. 7 shows the voxel-labeling uncertainty on the same representative subject. In general, the uncertainty of the brain extraction is very low, and the relatively higher uncertainty regions concentrate around the edges of the brain. The uncertainty map of each subject was also transformed, averaged, collapsed and displayed in the same manner as the averaged absolute error map to calculate and show the averaged uncertainty map in the NMT space (Fig. 8). Fig. 8 illustrates the systematic uncertainty distribution in the 3D volume over all the subjects. Overall, the uncertainty is very low, and the relative high uncertainty area is at boundary of the brain close to the brain stem. The fully connected 3D CRF successfully helped correct the results from Bayesian SegNet around this area as shown is Fig. 5 and 6.

In terms of processing time, the prediction for a single subject in the test stage by Bayesian SegNet with Nvidia GTX980Ti GPU is around 40 seconds, and after this the fully connected 3D CRF with an Intel Xeon(R) E5-2620 v4 CPU (8 cores, 16 threads @2.10GHz) is approximately 80 seconds for 5 iterations. Thus, the total time for one prediction is approximately 2 minutes, which is comparable to other edge detecting based methods. However, the template based method, NMT, is very time consuming. It costs around 5 hours even with an OpenMP version AFNI and 2 Intel Xeon(R) E5-2620 v4 CPUs to do the registration for one subject.

## 3.3 Uncertainty of the Bayesian SegNet

Uncertainty maps were also generated by Bayesian SegNet trained with different numbers of subjects to study the effect of training set size on the uncertainty. The uncertainty maps of the representative subject are shown in Fig 9, from which it can be seen that as the training set size decreases, the uncertainty increases, especially at the boundaries of the frontal lobe and behind the eyes. The total uncertainty defined as $\sigma_{tot} = \sqrt{\sum_i \sigma_i^2}$ ($i$ is the voxel index) was also calculated for each subject, and the total uncertainties for the 50 subjects in the testing set generated by different training set sizes are shown in the boxplot in Fig. 10. In Fig. 10, an increase in the total uncertainty can be seen as the training set decreases, and the total uncertainty of every subject tends to deviate more from one to another as the training set size decreases.

The relationship between training label consistency and prediction uncertainty was also studied. Fig. 11 shows the uncertainty maps of a representative subject 007 generated by Bayesian SegNet trained with manual labels and labels generated by AFNI+. As the number of AFNI+ labels increases in the 50-subject training set, the uncertainty generated by Bayesian SegNet also increases, especially in the frontal lobe and regions behind the eyes where the AFNI+ labels mismatch the corresponding manual labels. Fig. 12 shows a boxplot of the total uncertainty in the ROI behind the eyes against different AFNI+ label numbers in the training set. As the number of AFNI+ labels in the training set increases, each tested subject's total uncertainty in the inconsistently labeled area also increases and tends to deviate more from one to another.

The inconsistency between the training set and the testing set can also cause erroneous results. The uncertainty generated can give a warning about this kind of inconsistency. Fig.

13 shows the brain extraction performance of the proposed method with the same parameters (trained with the 50 subjects in fold 1 and using the same 3D CRF parameters) on 5 different testing sets. The original data in fold 2 are most consistent with the training data, and this set had the best performance. When the fold 2 data were rotated by increasingly larger amounts, the inconsistency of them against the training set was larger, and the brain extraction results were worse. Since the data collected on the older scanner were also slightly inconsistent with the training data due to slight contrast differences and age and gender differences between different subject groups, the results were also slightly worse. Fig. 14 shows the corresponding uncertainty behavior. When the testing set is inconsistent with the training set, the total uncertainty is higher. The more inconsistency there is, the more the total uncertainty increases.

## 4 Discussion

A new fully-automated brain extraction method is proposed as a combination of deep probabilistic neural network and fully connected 3D conditional random field, for the challenging task of brain extraction in nonhuman primates. The brain extraction results of the 100-subject dataset suggest that the proposed method can achieve higher accuracy and superior performance compared to state-of-the-art methods, as is measured by many different metrics. In addition, the proposed method is also highly time-efficient for a single prediction of a couple of minutes, with the facilitation of parallel computation.

The difficulties of nonhuman primate brain extraction are mainly due to the unique anatomical structures, especially the adipose tissue behind the eyes, the sharp curvature of the frontal lobe and generally more muscular and bony structures (Rohlfing et al., 2012a). For the competing nondeep-learning based methods, the brain extraction results are in good agreement with previous published studies (Wang et al., 2014). Edge detection methods and gradient based methods, like BSE and AFNI, can easily fail on the adipose tissue behind the eyes because of their proximity to the brain and high contrast to surrounding tissues (Iglesias et al., 2011; Wang et al., 2014). Algorithms involving surface expansion or deformable techniques, like BET and HWA, usually reach a result that either misses the frontal lobe, or includes areas out of the brain, because of the sharp curvature of the frontal lobe (Fennema-Notestine et al., 2006; Shattuck et al., 2001). Template registration based methods, like ROBEX and NMT, highly depend on the accuracy of registration. Since every subject must have its own anatomical specificity, which can be highly unique, given significant differences in age, gender and health conditions, the registration often has flaws. (Roy et al., 2017).

In the deep learning based methods, BSegNet is better than SegNet due to the involvement of Monte Carlo dropout training and testing (Kendall et al., 2015b), and BSegNetCRF is better than BSegNet since fully connected 3D CRF makes it possible to refine the probability results in a fully 3D context. BSegNetCRF, LiviaNet and VNet are all 3D methods using different strategies to take the 3D context into consideration. BSegNetCRF refines the results from a 2D neural network with 3D CRF; LiviaNet unstacks the original 3D volumes into small 3D patches for the 3D network; VNet downsamples the original 3D volumes and processes the whole downsampled 3D volume with the 3D network. The results show that BSegNetCRF outperforms LiviaNet and VNet in this application, and there could be multiple possible reasons for this. Directly processing large 3D images on current GPUs is very challenging due to the limit of GPU memory, so the design of a 3D network has to be relatively light and shallow to reduce the memory request from the network parameters. Even so, to process the 3D input, either a patch-based strategy or downsampling still need to be used. A 3D Patch-based strategy reduces the network's receptive field, so it is usually used for segmenting small structures. For regions as large as nonhuman primate brains, small regional errors can be caused. Meanwhile, downsampling can affect the results directly. LiviaNet and VNet also use different network designs, such as number of layers, stride and loss function, and these differences could also be the potential reasons for the performance differences in this application.

Overall, compared with nondeep-learning based methods, deep learning uses the training process to capture the features of the dataset with the help of a subset of manually labeled brains as the prior knowledge. This gives deep learning the ability to segment very complex structures. For brain extraction, it can exclude the complex ventricle structures within the brain in a manner as the manual labels are defined in the training stage (Kleesiek et al., 2016). Meanwhile, deep learning also offers more flexibility in brain extraction since one can define the brain region as preferred by the training labels, for example, including or excluding certain parts of the brain, like the brain stem or cerebellum.

As a convolutional encoder-decoder network, Bayesian SegNet reaches a balance of being both deep and light, which makes it a powerful tool in brain extraction. For being deep, it has 13 convolutional layers, 13 deconvolutional layers and 26 corresponding ReLU layers, which makes it deep enough to extract high level features with a considerable receptive field, while possessing sufficient nonlinearity to build the transformation from the original images to the brain extraction labels (Badrinarayanan et al., 2015a).

However, usually a deep neural network suffers from an enormous number of parameters to train, which has a high cost in terms of GPU memory, and time in training and predicting, and can also result in overfitting. In terms of also being light, Bayesian SegNet elegantly uses the pooling indices in the maximum pooling layers to perform the nonlinear upsampling in the corresponding upsampling layers. This eliminates the need for training in all the upsampling steps and makes the upsampled maps sparse. Moreover, it also uses small convolutional filters and no fully connected layers. All these features make it small in terms of the number of trainable parameters and efficient in terms of both the memory cost and computational time (Badrinarayanan et al., 2015b; Liu et al., 2017).

Being different from other deep learning based neural networks applied to brain extraction, Bayesian SegNet is a probabilistic neural network, so it has the ability to provide the uncertainty of the network on each prediction, as well as predict accurate labels for all pixels (Kendall et al., 2015a). It is important for a predictive system to generate model uncertainty as a part of the output, since meaningful uncertainty measurement is important for decision-making, especially biomedical applications where accuracy is extremely important. To replace any manual procedures with deep learning based methods means the conventional ground truth is not available any more in a real prediction, so an output including the confidence of the model on each specific case becomes very important. The uncertainty offered by Bayesian SegNet meets this very need. In a routine procedure implemented by Bayesian SegNet, every output uncertainty will be checked automatically against an empirical threshold to determine whether the result can be trusted or human intervention should be started. In our exploration of the uncertainty generated by Bayesian SegNet, we demonstrated that the behavior of the uncertainty generated by Monte Carlo dropout sampling matches our expectation very well. The uncertainty tends to increase and deviate more from subject to subject, as the size of the training set decreases, the inconsistency of training labels increases, or the inconsistency between the training set and testing set increases (Fig. 9-12). Considering each training process takes about 12 hours, we only studied a few training set sizes and mismatching label numbers. In addition, although the training process always drives the loss to converge, the training procedure itself is a stochastic process. A more thorough study of the behavior of the uncertainty, and to what extent the training process affects the uncertainty, are future research topics. In terms of the inconsistency between training and testing sets, there are several possible solutions to improve the robustness of the method. One is to make the training set more like the testing test. For example, to predict randomly rotated data, the training set can also be randomly rotated before training, or all the training and testing data can be aligned to a template before training and testing to make the model more robust. The other solution is to use transfer learning (Pan and Yang, 2010). To predict the data from another site or using another pulse sequence, the trained network can be further finetuned with a small subset of this kind of inconsistent data before the real prediction.

The combination of fully connected 3D CRF takes the probability maps from Bayesian SegNet's 2D predictions and moves forward to predictions in a fully 3D context. Because of the limitation of current GPU memory, and the huge data size of brain images, it is currently challenging to make fully 3D predictions through deep learning on a single GPU (Wachinger et al., 2017). Thus, we chose the 2D neural network to meet the GPU memory limitation without compromising the network performance, while making the whole 2D slice available for training and predicting. Then, we involve fully connected 3D CRF to implement a complete 3D prediction taking into account all the information from the entire original brain volume. Results shown from Fig. 3, 4 and 6 demonstrate the improvement made by the combination of this fully 3D process to the deep learning alone method. In addition, there can be errors in the direct results from deep learning based methods due to the imperfection and inconsistency of manual labeling. Fully-connected 3D CRF can fix these errors to some extent by taking into account the contrast and distance of all the connections in the original 3D image (shown in Fig. 5 and 6). The parameters of fully connected 3D CRF were empirically selected. To achieve the optimal performance, the weights for the result of deep learning, image intensity and voxel distance need to reach a balance. If the result of deep learning is over weighed, the effect of CRF's refinement won't be realized, and the result will be similar to that from deep learning. If the image intensity or voxel distance is over weighed, then the deep learning portion will be underemphasized, and the result could be worse than that from deep learning. As each of the parameters changes, the Dice coefficient will change gradually.

There are also some limitations in our study. To simulate the typical parameter-selecting procedure, we didn't carry out subject specific parameter selection, nor did we perform a grid search in the parameter space. It is possible that the performance of these methods can be improved with these strategies, however, they are very time consuming, and thus, impractical (Kleesiek et al., 2016). Another limitation is that we only studied the periadolescent rhesus monkeys. The structure of a nonhuman primate's skull and the amount of muscle tissue change dramatically across development, and infant monkeys are being used more and more in neuroscience studies (Kourtzi et al., 2006; Livingstone et al., 2017). Currently we are collecting brain MR images of rhesus monkeys across the whole age spectrum to test our method and investigate how transfer learning can be applied across different age groups. Future study also includes the possibility of combining the third-dimensional information (Xu et al., 2017) and image noise information (Kendall and Gal, 2017) of MRI brain volumes into the network. These approaches may further improve the performance of brain extraction.

## 5 Conclusion

In conclusion, we proposed and evaluated a new fully-automated brain extraction method integrating Bayesian SegNet and fully connected 3D CRF for nonhuman primate MRI brain images. Being different from previous designs, our approach is not only able to generate accurate and rapid brain extraction in a fully 3D context, but it also involves a probabilistic convolutional neural network that can output the uncertainty of the network on each prediction. This can greatly facilitate current large-scale MRI based neuroscience, neuroimaging and psychiatry studies on nonhuman primates.

## Supplementary Material

Refer to Web version on PubMed Central for supplementary material.

## Acknowledgments

The authors gratefully acknowledge the technical expertise of Ms. Maria Jesson and Dr. Andrew Fox (UC-Davis), and the assistance of the staffs at the Harlow Center for Biological Psychology, the Lane Neuroimaging Laboratory at the Health Emotions Research Institute, Waisman Laboratory for Brain Imaging and Behavior, and the Wisconsin National Primate Research Center. This work was supported by grants from the National Institutes of Health: P51-OD011106; R01-MH046729; R01-MH081884; P50-MH100031. The content is solely the responsibility of the authors and does not necessarily represent the official views of the National Institutes of Health.

## References

Badrinarayanan V, Handa A, Cipolla R. SegNet: A Deep Convolutional Encoder-Decoder Architecture for Robust Semantic Pixel-Wise Labelling. ArXiv150507293 Cs. 2015a

Badrinarayanan V, Kendall A, Cipolla R. SegNet: A Deep Convolutional Encoder-Decoder Architecture for Image Segmentation. ArXiv151100561 Cs. 2015b

Badrinarayanan V, Kendall A, Cipolla R. SegNet: A Deep Convolutional Encoder-Decoder Architecture for Image Segmentation. ArXiv151100561 Cs. 2015c

Baldwin RM, Zea-Ponce Y, Zoghbi sami S, Laurelle M, Al-Tikriti MS, Sybirska EH, Malison RT, Neumeyer JL, Milius RA, Wang S, Stabin M, Smith EO, Charney DS, Hoffer PB, Innis RB. Evaluation of the monoamine uptake site ligand [131I]methyl 3β-(4-Iodophenyl)-tropane-2β-carboxylate ([123I]β-CIT) in non-human primates: Pharmacokinetics, biodistribution and SPECT

brain imaging coregistered with MRI. Nucl Med Biol. 1993; 20:597–606. DOI: 10.1016/0969-8051(93)90028-S [PubMed: 8358345]

Bernal J, Kushibar K, Asfaw DS, Valverde S, Oliver A, Martí R, Lladó X. Deep convolutional neural networks for brain image analysis on magnetic resonance imaging: a review. ArXiv171203747 Cs. 2017

Birn RM, Shackman AJ, Oler JA, Williams LE, McFarlin DR, Rogers GM, Shelton SE, Alexander AL, Pine DS, Slattery MJ, Davidson RJ, Fox AS, Kalin NH. Evolutionarily conserved prefrontal-amygdalar dysfunction in early-life anxiety. Mol Psychiatry. 2014; 19:915–922. DOI: 10.1038/mp.2014.46 [PubMed: 24863147]

Chen H, Dou Q, Yu L, Qin J, Heng P-A. VoxResNet: Deep voxelwise residual networks for brain segmentation from 3D MR images. NeuroImage. 2017.

Çiçek Ö, Abdulkadir A, Lienkamp SS, Brox T, Ronneberger O. 3D U-Net: Learning Dense Volumetric Segmentation from Sparse Annotation. ArXiv160606650 Cs. 2016

Coates A, Huval B, Wang T, Wu D, Catanzaro B, Andrew N. Deep learning with COTS HPC systems. International Conference on Machine Learning. 2013:1337–1345.

Craddock RC, Bellec P, Jbabdi S. Neuroimage special issue on brain segmentation and parcellation - Editorial. NeuroImage. 2017.

Dice LR. Measures of the amount of ecologic association between species. Ecology. 1945; 26:297–302.

Dolz J, Desrosiers C, Ben Ayed I. 3D fully convolutional networks for subcortical segmentation in MRI: A large-scale study. NeuroImage. 2017.

Everingham M, Eslami SMA, Gool LV, Williams CKI, Winn J, Zisserman A. The Pascal Visual Object Classes Challenge: A Retrospective. Int J Comput Vis. 2015; 111:98–136. DOI: 10.1007/s11263-014-0733-5

Fennema-Notestine C, Ozyurt IB, Clark CP, Morris S, Bischoff-Grethe A, Bondi MW, Jernigan TL, Fischl B, Segonne F, Shattuck DW, Leahy RM, Rex DE, Toga AW, Zou KH, BIRN M, Brown GG. Quantitative Evaluation of Automated Skull-Stripping Methods Applied to Contemporary and Legacy Images: Effects of Diagnosis, Bias Correction, and Slice Location. Hum Brain Mapp. 2006; 27:99–113. DOI: 10.1002/hbm.20161 [PubMed: 15986433]

Fox AS, Kalin NH. A translational neuroscience approach to understanding the development of social anxiety disorder and its pathophysiology. Am J Psychiatry. 2014; 171:1162–1173. DOI: 10.1176/appi.ajp.2014.14040449 [PubMed: 25157566]

Fox AS, Oler JA, Shackman AJ, Shelton SE, Raveendran M, McKay DR, Converse AK, Alexander A, Davidson RJ, Blangero J, Rogers J, Kalin NH. Intergenerational neural mediators of early-life anxious temperament. Proc Natl Acad Sci U S A. 2015a; 112:9118–9122. DOI: 10.1073/pnas.1508593112 [PubMed: 26150480]

Fox AS, Oler JA, Shackman AJ, Shelton SE, Raveendran M, McKay DR, Converse AK, Alexander A, Davidson RJ, Blangero J, Rogers J, Kalin NH. Intergenerational neural mediators of early-life anxious temperament. Proc Natl Acad Sci U S A. 2015b; 112:9118–9122. DOI: 10.1073/pnas.1508593112 [PubMed: 26150480]

Gal Y, Ghahramani Z. Bayesian Convolutional Neural Networks with Bernoulli Approximate Variational Inference. ArXiv150602158 Cs Stat. 2015

Geremia E, Clatz O, Menze BH, Konukoglu E, Criminisi A, Ayache N. Spatial decision forests for MS lesion segmentation in multi-channel magnetic resonance images. NeuroImage. 2011; 57:378–390. [PubMed: 21497655]

Gerig G, Jomier M, Chakos M. Valmet: A New Validation Tool for Assessing and Improving 3D Object Segmentation. Medical Image Computing and Computer-Assisted Intervention – MICCAI 2001; Presented at the International Conference on Medical Image Computing and Computer-Assisted Intervention; Berlin, Heidelberg: Springer; 2001. 516–523.

He X, Zemel RS, Carreira-Perpinan MA. Multiscale conditional random fields for image labeling. Proceedings of the 2004 IEEE Computer Society Conference on Computer Vision and Pattern Recognition, 2004: CVPR 2004; Presented at the Proceedings of the 2004 IEEE Computer Society Conference on Computer Vision and Pattern Recognition, 2004. CVPR 2004; 2004. II-695–II-702.

Huttenlocher DP, Klanderman GA, Rucklidge WJ. Comparing images using the Hausdorff distance. IEEE Trans Pattern Anal Mach Intell. 1993; 15:850–863.

Iglesias JE, Liu CY, Thompson PM, Tu Z. Robust Brain Extraction Across Datasets and Comparison With Publicly Available Methods. IEEE Trans Med Imaging. 2011; 30:1617–1634. DOI: 10.1109/TMI.2011.2138152 [PubMed: 21880566]

Ioffe S, Szegedy C. Batch Normalization: Accelerating Deep Network Training by Reducing Internal Covariate Shift. ArXiv150203167 Cs. 2015

Kalin NH, Shelton SE, Davidson RJ. Role of the Primate Orbitofrontal Cortex in Mediating Anxious Temperament. Biol Psychiatry. 2007; 62:1134–1139. DOI: 10.1016/j.biopsych.2007.04.004 [PubMed: 17643397]

Kamnitsas K, Ledig C, Newcombe VFJ, Simpson JP, Kane AD, Menon DK, Rueckert D, Glocker B. Efficient multi-scale 3D CNN with fully connected CRF for accurate brain lesion segmentation. Med Image Anal. 2017; 36:61–78. DOI: 10.1016/j.media.2016.10.004 [PubMed: 27865153]

Kendall A, Badrinarayanan V, Cipolla R. Bayesian SegNet: Model Uncertainty in Deep Convolutional Encoder-Decoder Architectures for Scene Understanding. ArXiv151102680 Cs. 2015a

Kendall A, Badrinarayanan V, Cipolla R. Bayesian SegNet: Model Uncertainty in Deep Convolutional Encoder-Decoder Architectures for Scene Understanding. ArXiv151102680 Cs. 2015b

Kendall A, Gal Y. What Uncertainties Do We Need in Bayesian Deep Learning for Computer Vision? ArXiv170304977 Cs. 2017

Kleesiek J, Urban G, Hubert A, Schwarz D, Maier-Hein K, Bendszus M, Biller A. Deep MRI brain extraction: A 3D convolutional neural network for skull stripping. NeuroImage. 2016; 129:460–469. DOI: 10.1016/j.neuroimage.2016.01.024 [PubMed: 26808333]

Kourtzi Z, Augath M, Logothetis NK, Movshon JA, Kiorpes L. Development of visually evoked cortical activity in infant macaque monkeys studied longitudinally with fMRI. Magn Reson Imaging. 2006; 24:359–366. DOI: 10.1016/j.mri.2005.12.025 [PubMed: 16677941]

Krähenbühl P, Koltun V. Efficient Inference in Fully Connected CRFs with Gaussian Edge Potentials. ArXiv12105644 Cs. 2012

Krizhevsky A, Sutskever I, Hinton GE. ImageNet Classification with Deep Convolutional Neural Networks. In: Pereira F, Burges CJC, Bottou L, Weinberger KQ, editorsAdvances in Neural Information Processing Systems 25. Curran Associates, Inc.; 2012a. 1097–1105.

Krizhevsky A, Sutskever I, Hinton GE. Imagenet classification with deep convolutional neural networks. Advances in Neural Information Processing Systems. 2012b:1097–1105.

LeCun Y, Bengio Y, Hinton G. Deep learning. Nature. 2015; 521:436–444. DOI: 10.1038/nature14539 [PubMed: 26017442]

Liu F, Zhou Z, Jang H, Samsonov A, Zhao G, Kijowski R. Deep convolutional neural network and 3D deformable approach for tissue segmentation in musculoskeletal magnetic resonance imaging. Magn Reson Med. 2017.

Livingstone MS, Vincent JL, Arcaro MJ, Srihasam K, Schade PF, Savage T. Development of the macaque face-patch system. Nat Commun. 2017; 8doi: 10.1038/ncomms14897

Long J, Shelhamer E, Darrell T. Fully Convolutional Networks for Semantic Segmentation. Presented at the Proceedings of the IEEE Conference on Computer Vision and Pattern Recognition; 2015. 3431–3440.

Milletari F, Navab N, Ahmadi S-A. V-Net: Fully Convolutional Neural Networks for Volumetric Medical Image Segmentation. ArXiv160604797 Cs. 2016

Nair V, Hinton GE. Rectified linear units improve restricted boltzmann machines. Proceedings of the 27th International Conference on Machine Learning (ICML-10); 2010. 807–814.

Oler JA, Fox AS, Shelton SE, Rogers J, Dyer TD, Davidson RJ, Shelledy W, Oakes TR, Blangero J, Kalin NH. Amygdalar and hippocampal substrates of anxious temperament differ in their heritability. Nature. 2010; 466:864–868. DOI: 10.1038/nature0928 [PubMed: 20703306]

Pan SJ, Yang Q. A Survey on Transfer Learning. IEEE Trans Knowl Data Eng. 2010; 22:1345–1359. DOI: 10.1109/TKDE.2009.191

Rohlfing T, Kroenke CD, Sullivan EV, Dubach MF, Bowden DM, Grant K, Pfefferbaum A. The INIA19 Template and NeuroMaps Atlas for Primate Brain Image Parcellation and Spatial Normalization. Front Neuroinformatics. 2012a; 6doi: 10.3389/fninf.2012.00027

Rohlfing T, Kroenke CD, Sullivan EV, Dubach MF, Bowden DM, Grant KA, Pfefferbaum A. The INIA19 Template and NeuroMaps Atlas for Primate Brain Image Parcellation and Spatial Normalization. Front Neuroinformatics. 2012b; 6doi: 10.3389/fninf.2012.00027

Ronneberger O, Fischer P, Brox T. U-Net: Convolutional Networks for Biomedical Image Segmentation. ArXiv150504597 Cs. 2015

Roy S, Butman JA, Pham DL. Robust skull stripping using multiple MR image contrasts insensitive to pathology. NeuroImage. 2017; 146:132–147. DOI: 10.1016/j.neuroimage.2016.11.017 [PubMed: 27864083]

Salehi SSM, Erdogmus D, Gholipour A. Auto-context Convolutional Neural Network (Auto-Net) for Brain Extraction in Magnetic Resonance Imaging; IEEE Trans Med Imaging PP. 2017. 1–1.

Schmidhuber J. Deep learning in neural networks: An overview. Neural Netw. 2015; 61:85–117. DOI: 10.1016/j.neunet.2014.09.003 [PubMed: 25462637]

Ségonne F, Dale AM, Busa E, Glessner M, Salat D, Hahn HK, Fischl B. A hybrid approach to the skull stripping problem in MRI. NeuroImage. 2004; 22:1060–1075. DOI: 10.1016/j.neuroimage.2004.03.032 [PubMed: 15219578]

Seidlitz J, Sponheim C, Glen D, Ye FQ, Saleem KS, Leopold DA, Ungerleider L, Messinger A. A population MRI brain template and analysis tools for the macaque. NeuroImage. 2017.

Shackman AJ, Fox AS, Oler JA, Shelton SE, Oakes TR, Davidson RJ, Kalin NH. Heightened extended amygdala metabolism following threat characterizes the early phenotypic risk to develop anxiety-related psychopathology. Mol Psychiatry. 2017; 22:724–732. DOI: 10.1038/mp.2016.132 [PubMed: 27573879]

Shattuck DW, Prasad G, Mirza M, Narr KL, Toga AW. Online resource for validation of brain segmentation methods. NeuroImage. 2009; 45:431–439. DOI: 10.1016/j.neuroimage.2008.10.066 [PubMed: 19073267]

Shattuck DW, Sandor-Leahy SR, Schaper KA, Rottenberg DA, Leahy RM. Magnetic Resonance Image Tissue Classification Using a Partial Volume Model. NeuroImage. 2001; 13:856–876. DOI: 10.1006/nimg.2000.0730 [PubMed: 11304082]

Shelhamer E, Long J, Darrell T. Fully Convolutional Networks for Semantic Segmentation. ArXiv160506211 Cs. 2016

Shotton J, Winn J, Rother C, Criminisi A. Textonboost for image understanding: Multi-class object recognition and segmentation by jointly modeling texture, layout, and context. Int J Comput Vis. 2009; 81:2–23.

Simonyan K, Zisserman A. Very Deep Convolutional Networks for Large-Scale Image Recognition. ArXiv14091556 Cs. 2014

Smith SM. Fast robust automated brain extraction. Hum Brain Mapp. 2002; 17:143–155. DOI: 10.1002/hbm.10062 [PubMed: 12391568]

Srivastava N, Hinton GE, Krizhevsky A, Sutskever I, Salakhutdinov R. Dropout: a simple way to prevent neural networks from overfitting. J Mach Learn Res. 2014; 15:1929–1958.

Taha AA, Hanbury A. Metrics for evaluating 3D medical image segmentation: analysis, selection, and tool. BMC Med Imaging. 2015; 15doi: 10.1186/s12880-015-0068-x

van der Kouwe AJW, Benner T, Salat DH, Fischl B. Brain morphometry with multiecho MPRAGE. NeuroImage. 2008; 40:559–569. DOI: 10.1016/j.neuroimage.2007.12.025 [PubMed: 18242102]

Wachinger C, Reuter M, Klein T. DeepNAT: Deep convolutional neural network for segmenting neuroanatomy. NeuroImage. 2017.

Wang Y, Jia H, Yap P-T, Cheng B, Wee C-Y, Guo L, Shen D. Groupwise Segmentation Improves Neuroimaging Classification Accuracy. In: Yap P-T, Liu T, Shen D, Westin C-F, Shen L, editorsProceedings; Multimodal Brain Image Analysis: Second International Workshop, MBIA 2012, Held in Conjunction with MICCAI 2012; Nice, France. October 1-5, 2012; Berlin, Heidelberg: Springer Berlin Heidelberg; 2012. 185–193.

Wang Y, Nie J, Yap P-T, Li G, Shi F, Geng X, Guo L, Shen D. Knowledge-Guided Robust MRI Brain Extraction for Diverse Large-Scale Neuroimaging Studies on Humans and Non-Human Primates. PLoS ONE. 2014; 9doi: 10.1371/journal.pone.0077810

Xu Y, Géraud T, Bloch I. From Neonatal to Adult Brain MR Image Segmentation in a Few Seconds Using 3D-Like Fully Convolutional Network and Transfer Learning. Proceedings of the 23rd IEEE International Conference on Image Processing (ICIP). Presented at the ICIP; 2017. 4417–4421.

Zhang D, Lu G. Review of shape representation and description techniques. Pattern Recognit. 2004; 37:1–19.

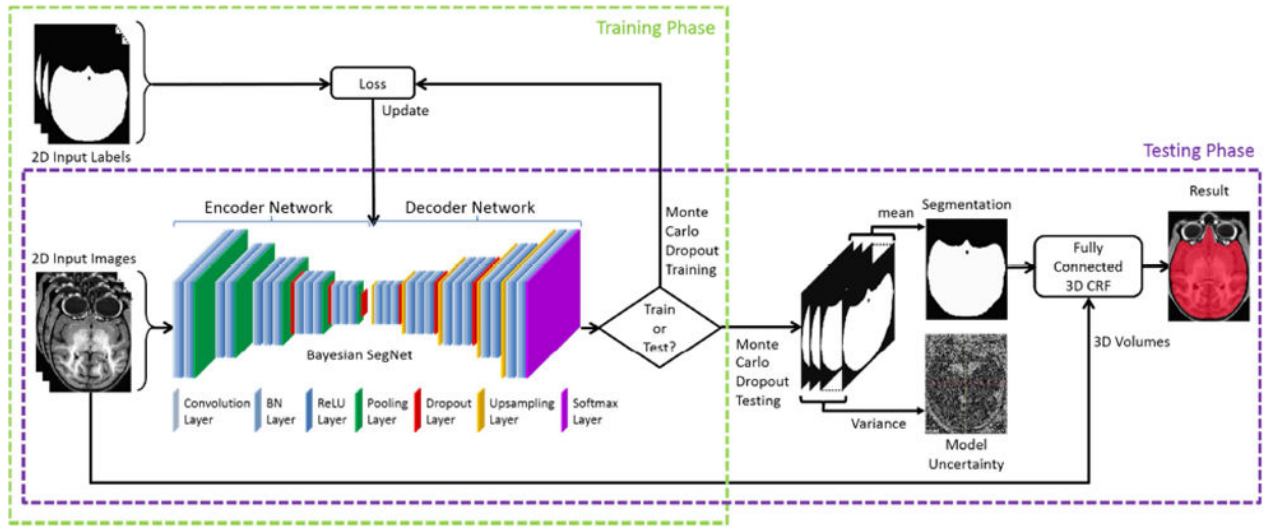

**Fig. 1.**
Work flow of the proposed brain extraction method, a combination of Bayesian SegNet and fully connected 3D CRF.

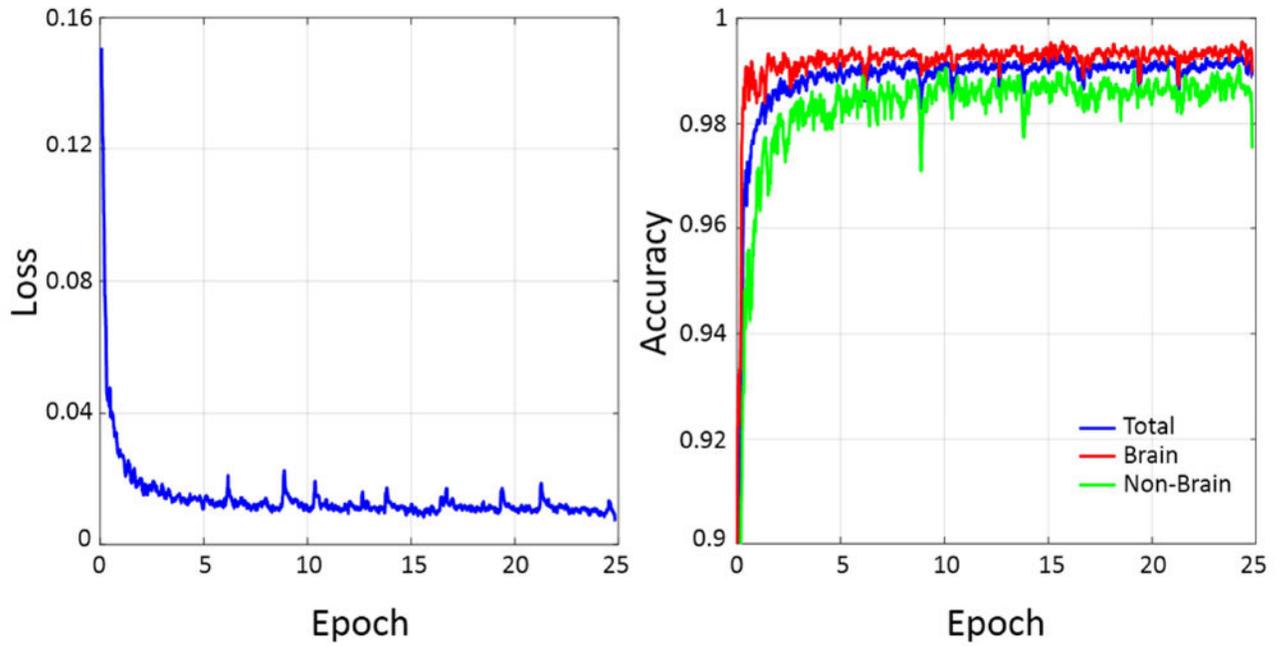

**Fig. 2.**
Loss and accuracy for Bayesian SegNet during training against epochs. Loss is the multinomial logistic loss between the output labels and the ground truth labels. Accuracy is defined as the ratio of correctly labeled pixels over total number of pixels for each category.

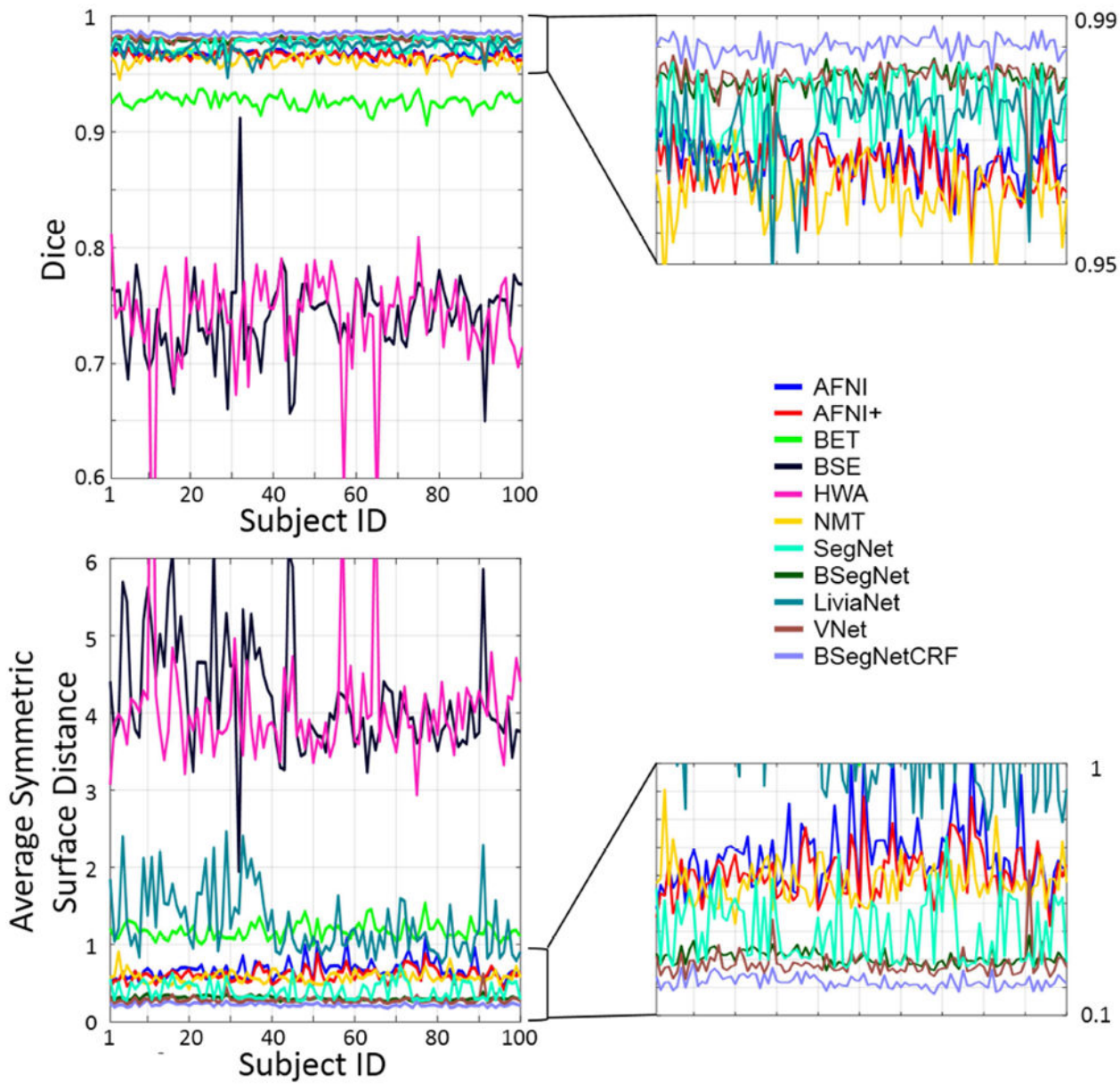

**Fig. 3.**
Evaluation scores on each subject from different brain extraction methods. Enlarged figures are on the right. Higher Dice coefficients and lower average symmetric surface distance indicate better agreement between the automatically-defined and manually-labeled (ground truth) brain masks. For all subjects, BSegNetCRF resulted in better brain extraction than all other methods tested.

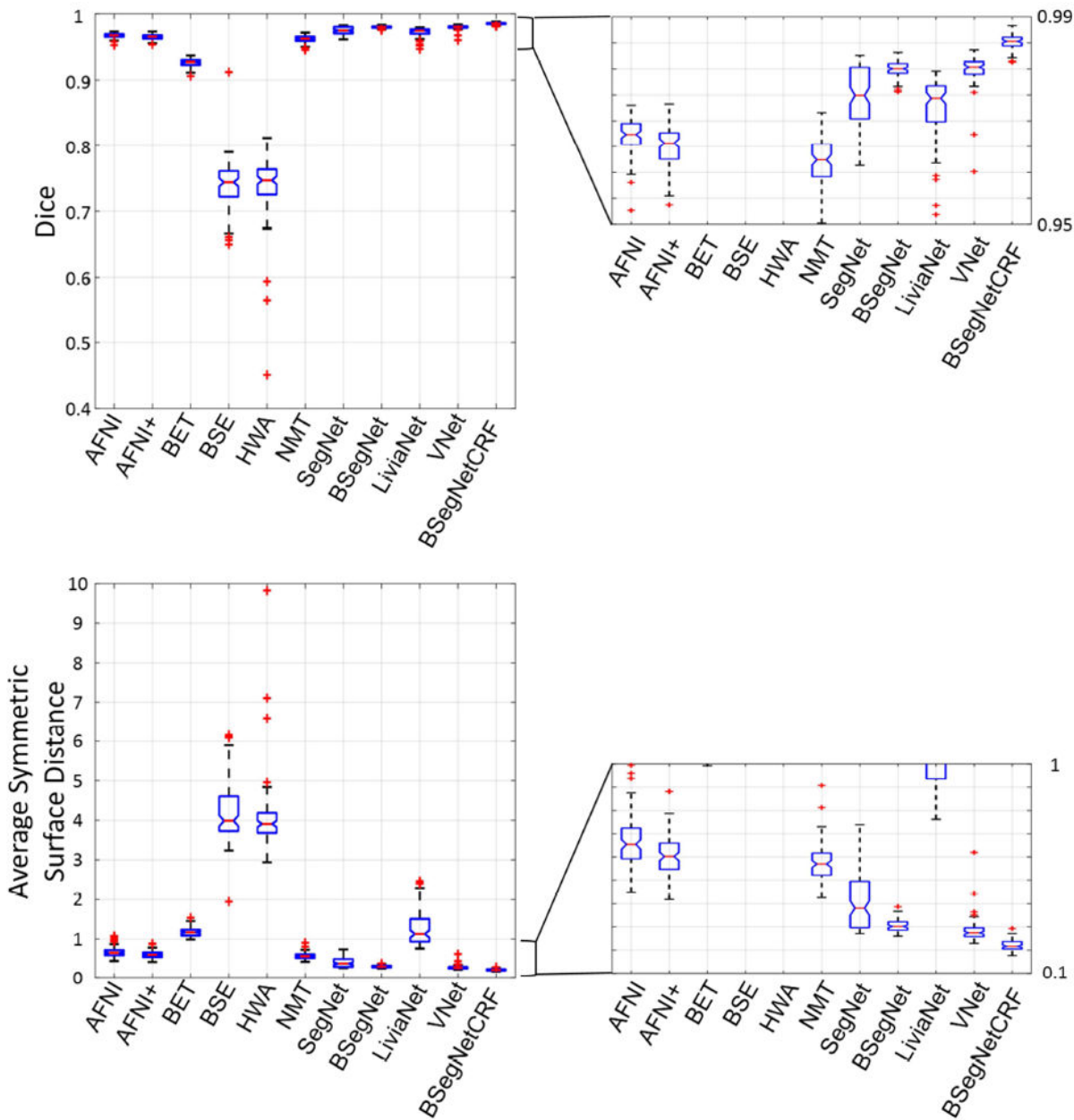

**Fig. 4.**
Evaluation scores in boxplots from different brain extraction methods. Enlarged figures are on the right. In the figure points are drawn as outliers with red '+' symbols, if they are greater than $q_3+1.5(q_3-q_1)$ or less than $q_1-1.5(q_3-q_1)$, where $q_1$ and $q_3$ are the first and third quartiles respectively.

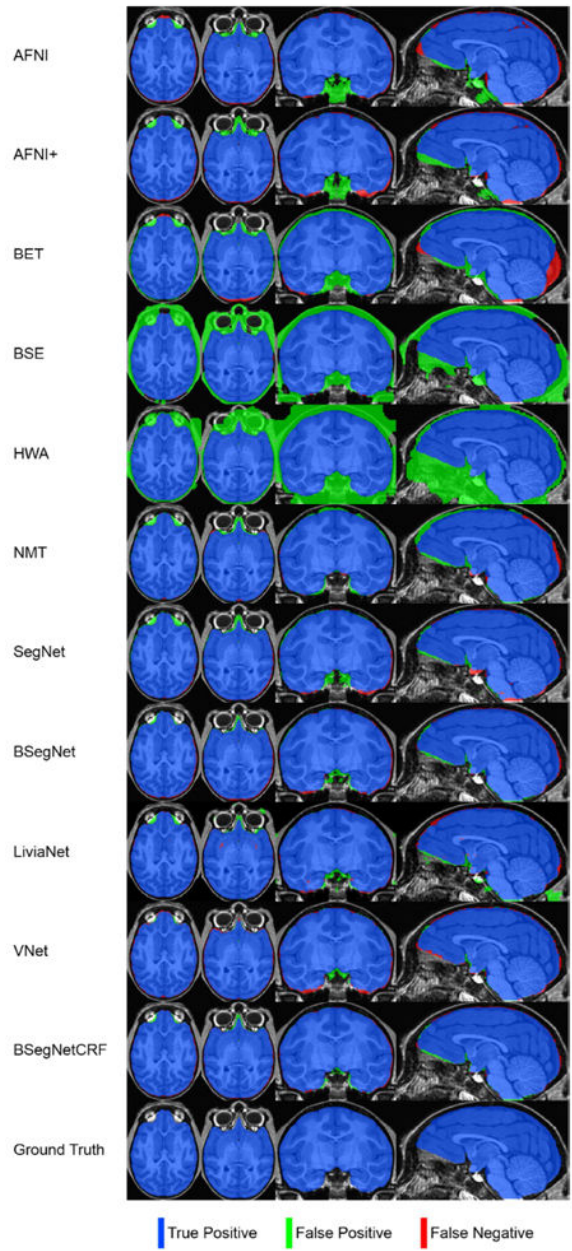

**Fig. 5.**
Comparison of the brain masks extracted by different methods on a typical subject: subject 007.

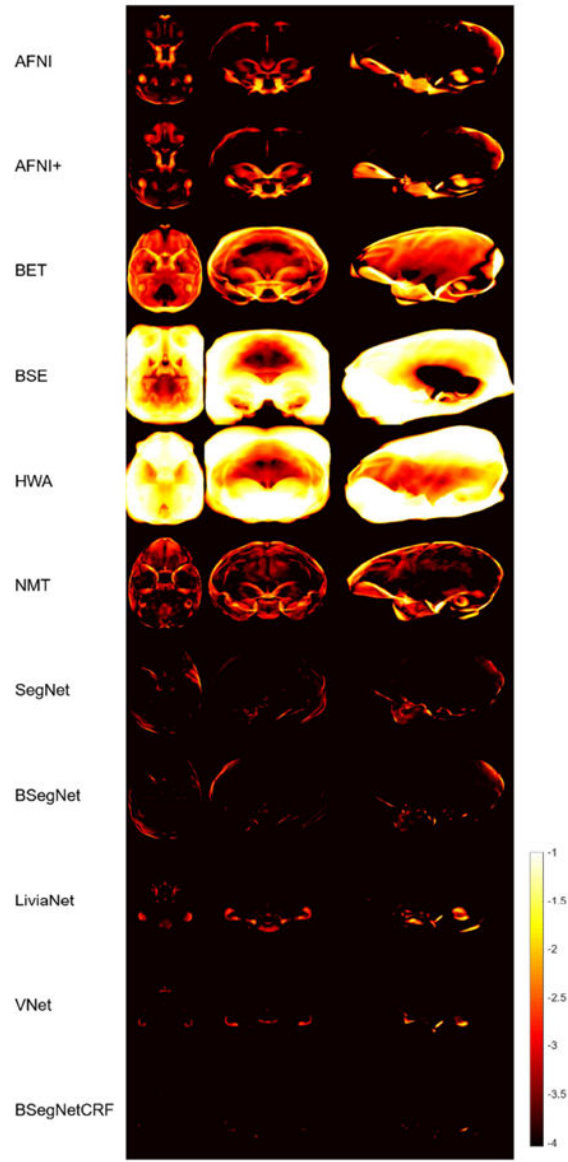

**Fig. 6.**
Averaged absolute error maps for compared methods. For display purposes, the natural logarithm of the averaged map collapsed (averaged) along each axis is shown.

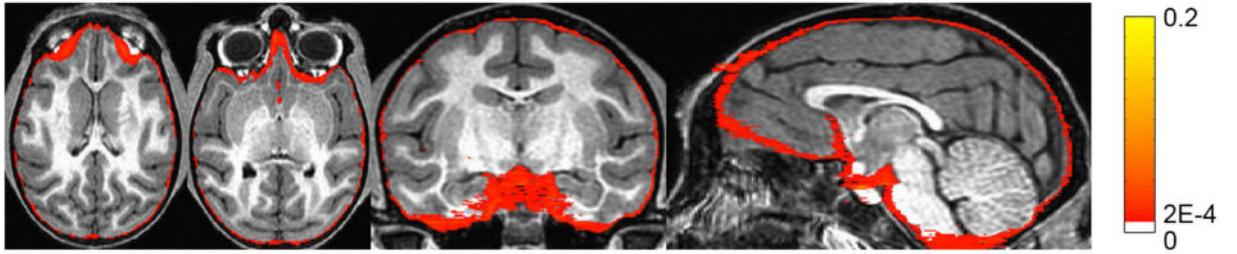

**Fig. 7.**
The uncertainty map given by Bayesian SegNet for subject 007.

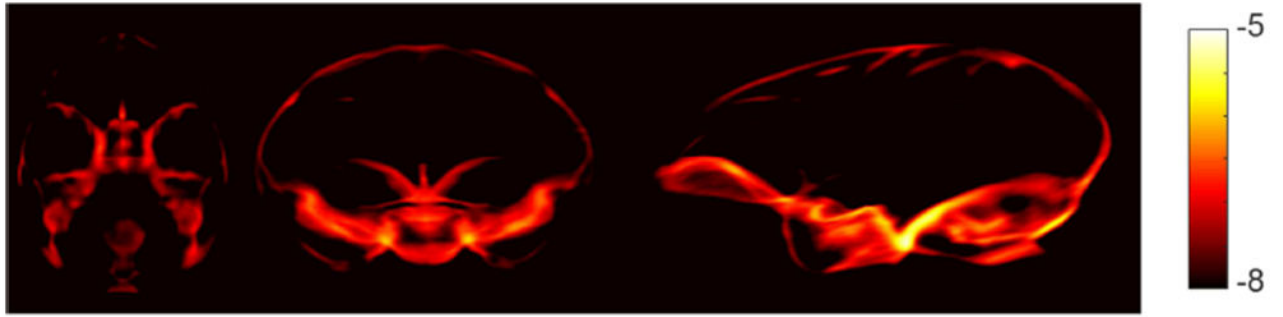

**Fig. 8.**
Averaged uncertainty maps from Bayesian SegNet. For display purposes, the natural logarithm of the averaged map collapsed (averaged) along each axis is shown.

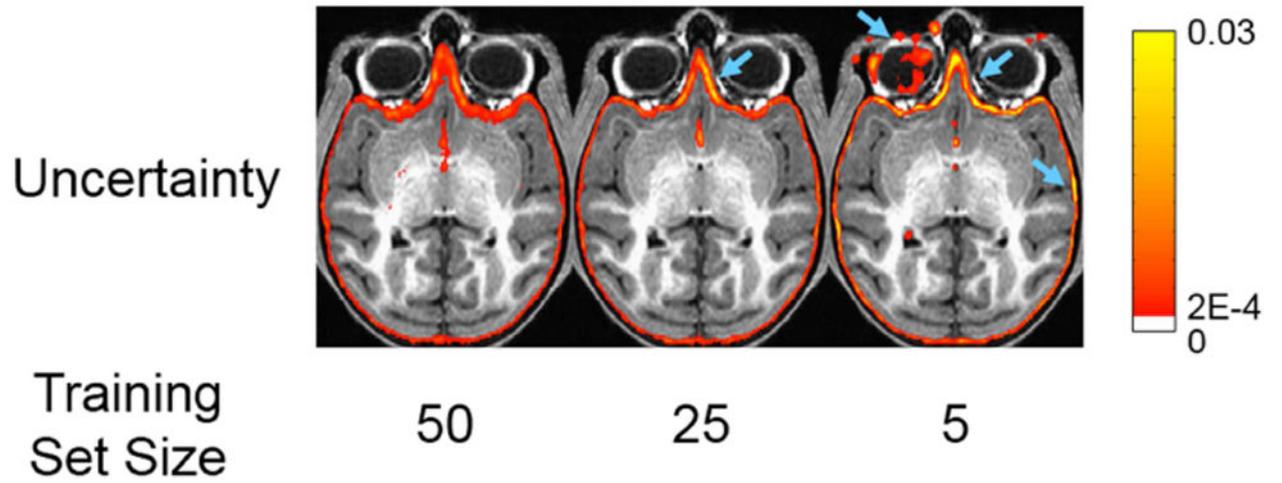

**Fig. 9.**
Uncertainty maps on subject 007 generated by Bayesian SegNet trained with different training set sizes. The blue arrows in the figure point out the regions with obvious uncertainty increase.

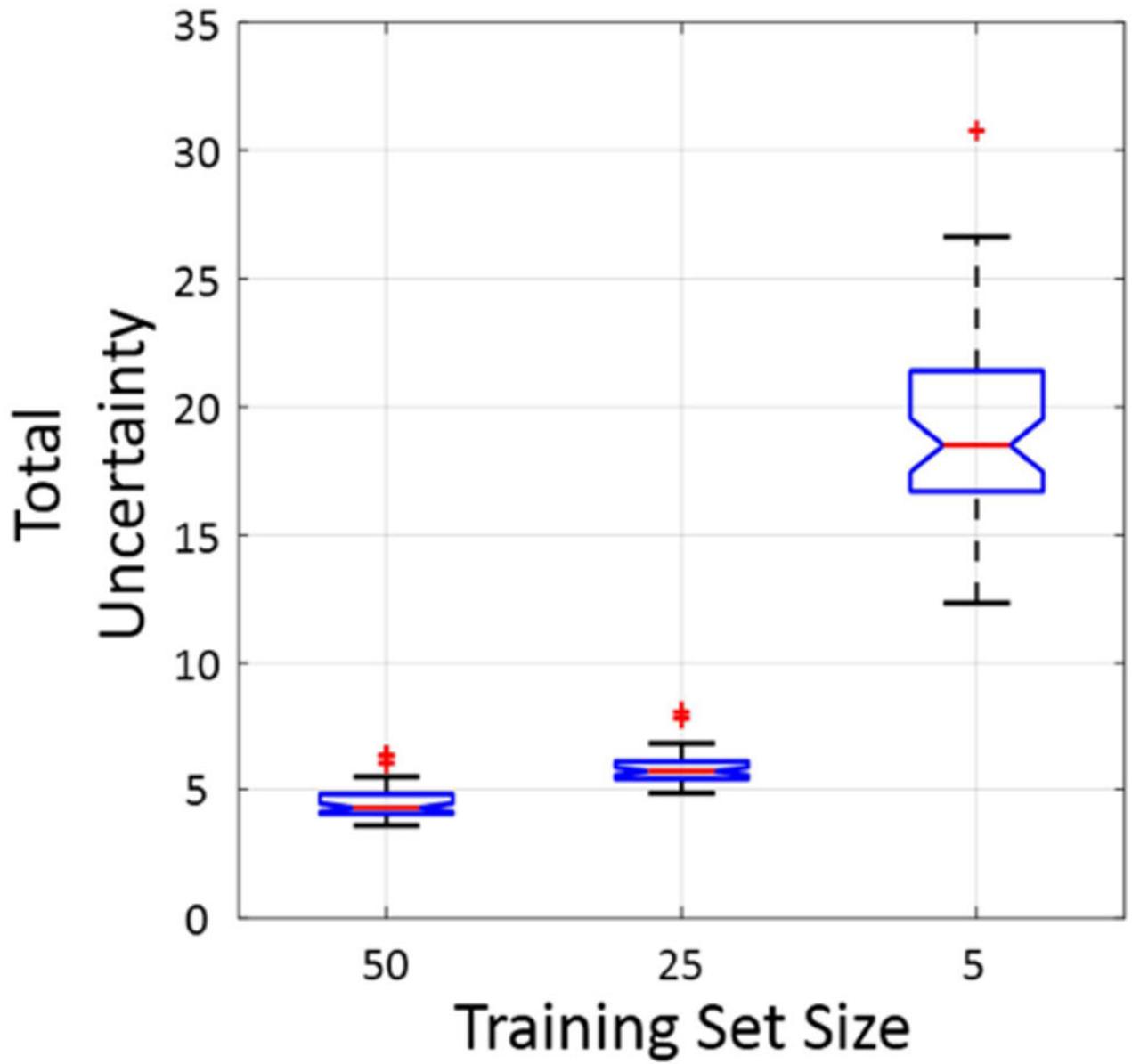

**Fig. 10.**
Total uncertainty in boxplots generated by Bayesian SegNet trained with different training set sizes. In the figure points are drawn as outliers with red '+' symbols, if they are greater than q3+1.5(q3-q1) or less than q1-1.5(q3-q1), where q1 and q3 are the first and third quartiles respectively.

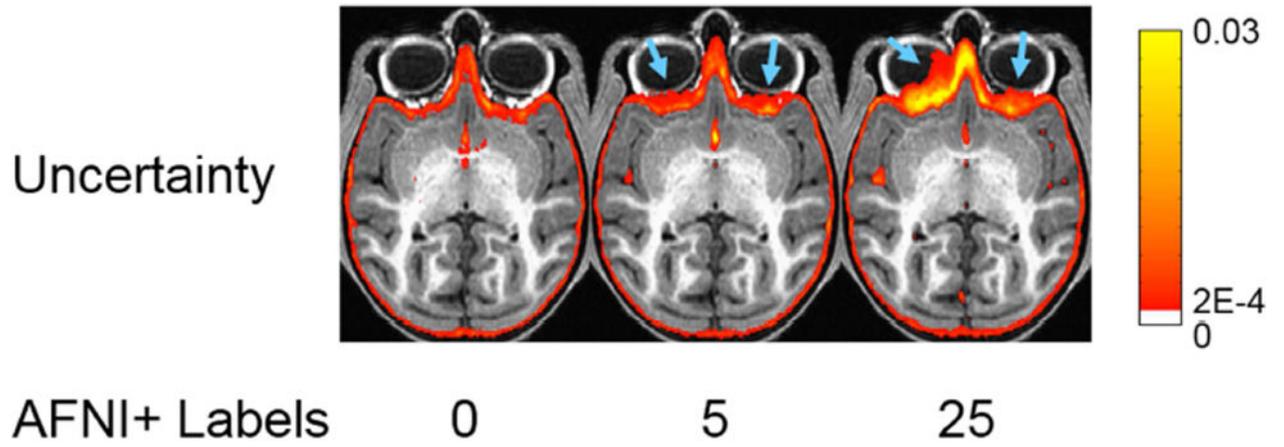

**Fig. 11.**
Uncertainty maps on subject 007 generated by Bayesian SegNet trained with 50 subjects in which different numbers of manual labels were replaced by labels generated by AFNI+ for the corresponding subjects. The blue arrows in the figure point out the regions with obvious uncertainty increase.

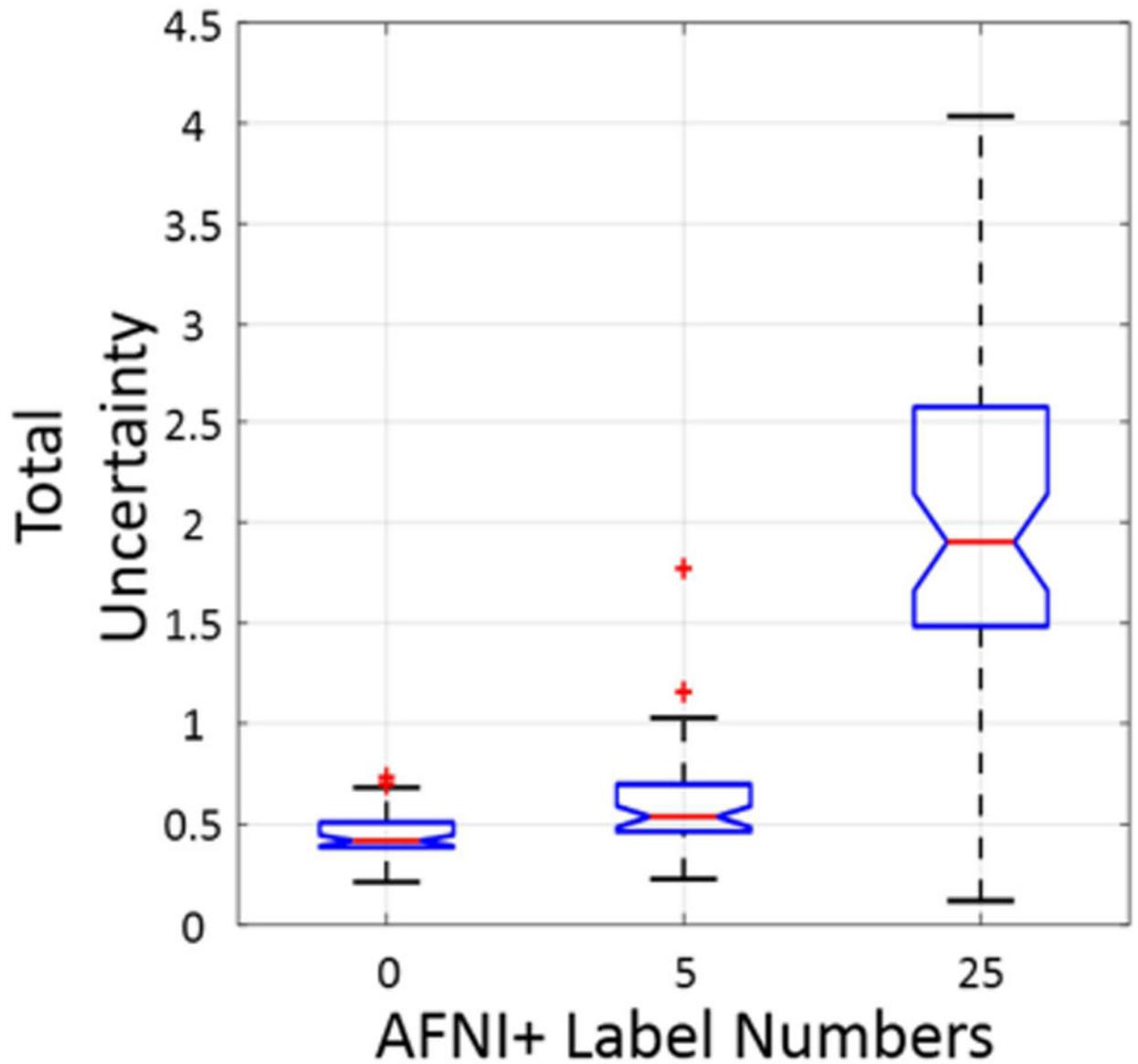

**Fig. 12.**
Total uncertainty of the ROI behind eyes in boxplots generated by Bayesian SegNet trained with 50 subjects in which different numbers of manual labels were replaced by labels generated by AFNI for the corresponding subjects. In the figure points are drawn as outliers with red '+' symbols, if they are greater than q3+1.5(q3-q1) or less than q1-1.5(q3-q1), where q1 and q3 are the first and third quartiles respectively.

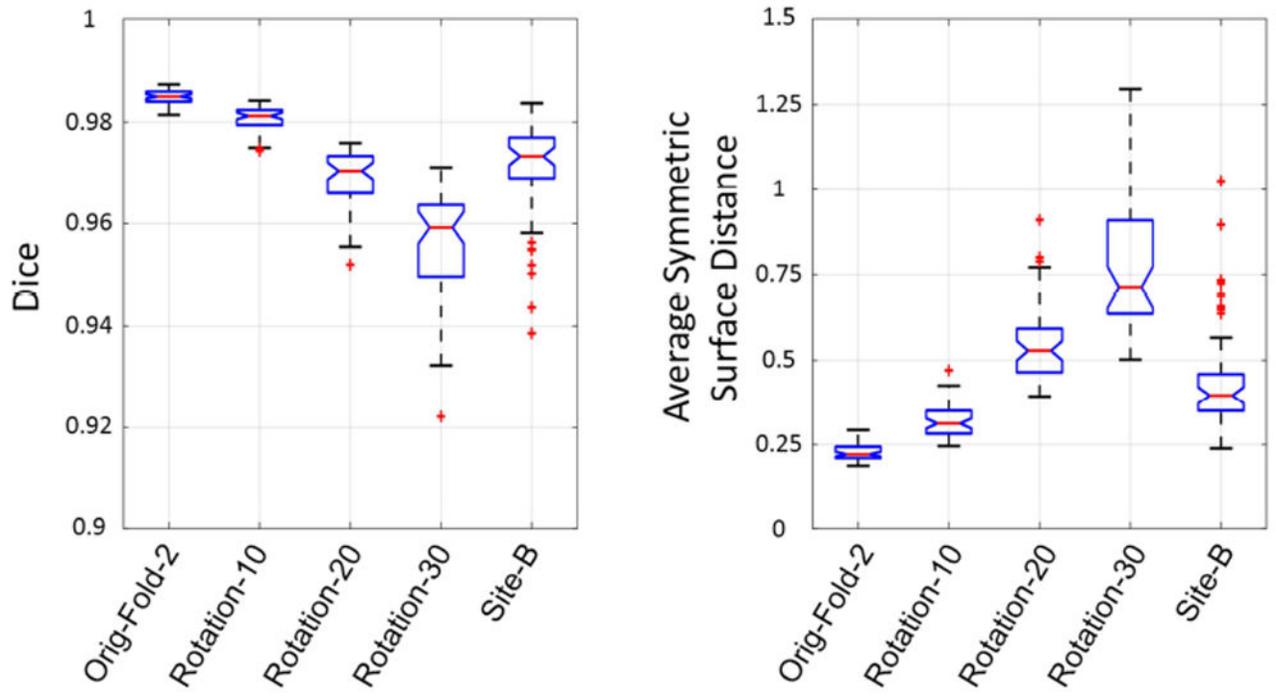

**Fig. 13.**
Evaluation scores in boxplots for the original fold 2 data, rotated fold 2 data and data from another site. In the figure points are drawn as outliers with red '+' symbols, if they are greater than q3+1.5(q3-q1) or less than q1-1.5(q3-q1), where q1 and q3 are the first and third quartiles respectively.

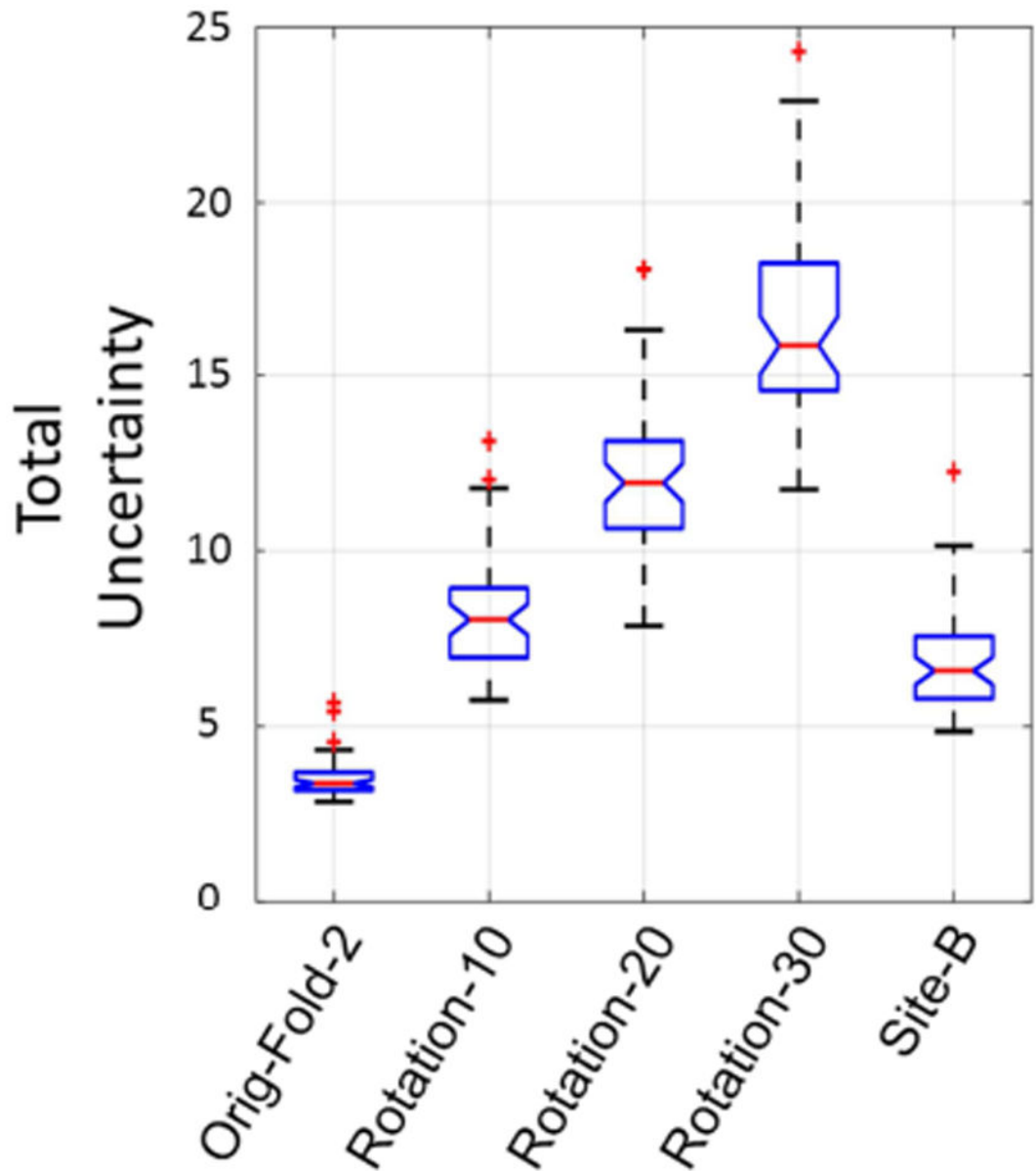

**Fig. 14.**
Total uncertainty generated by Bayesian SegNet for the original fold 2 data, rotated fold 2 data and data from another site in boxplots. In the figure points are drawn as outliers with red '+' symbols, if they are greater than q3+1.5(q3-q1) or less than q1-1.5(q3-q1), where q1 and q3 are the first and third quartiles respectively.

**Table 1**

Parameters studied and values used for the competing methods.

| Method | Parameter | Description | Range | Optimal value |
|--------|-----------|-------------|-------|---------------|
| 3dSkullStrip | -push_to_edge | Push to edge aggressively | w/ or w/o | w/ |
| | -monkey | Brain of a monkey | w/ or w/o | w/ |
| | -shrink_fac | Brain VS non-brain intensity threshold | 0~1 | 0.5 for AFNI |
| | | | | 0.4 for AFNI+ |
| HWA | -less | Shrink the surface | w/ or w/o | w/o |
| | -more | Expand the surface | w/ or w/o | w/o |
| | -atlas | Use the atlas information | w/ or w/o | w/ |
| BET | -f | Fractional intensity threshold | 0.1~0.9 | 0.3 |
| | -g | Vertical gradient | -1~1 | -0.5 |
| | -r | Head radius | 30~50 | 35 |
| BSE | -d | Diffusion constant | 5~35 | 25 |
| | -s | Edge detection constant | 0.10~0.80 | 0.69 |
| | -p | Dilate final mask | w/ or w/o | w/o |

**Table 2**

Mean and standard deviation of Dice coefficient and average symmetric surface distance (ASSD) for all 100 subjects. The best result is in bold font.

| Method | Dice | ASSD/mm |
|---|---|---|
| AFNI | 0.967 (±0.003) | 0.670 (±0.123) |
| AFNI+ | 0.965 (±0.004) | 0.609 (±0.088) |
| BET | 0.926 (±0.006) | 1.175 (±0.105) |
| BSE | 0.740 (±0.034) | 4.200 (±0.738) |
| HWA | 0.739 (±0.046) | 4.046 (±0.812) |
| NMT | 0.962 (±0.005) | 0.578 (±0.075) |
| SegNet | 0.975 (±0.006) | 0.404 (±0.116) |
| BSegNet | 0.980 (±0.002) | 0.306 (±0.026) |
| LiviaNet | 0.972 (±0.006) | 1.271 (±0.431) |
| VNet | 0.980 (±0.003) | 0.283 (±0.046) |
| BSegNetCRF | **0.985 (±0.002)** | **0.220 (±0.023)** |